\documentclass[11pt,a4paper]{article}
\usepackage[utf8]{inputenc}
\usepackage{jheppub}
\usepackage{slashed}
\usepackage{amsmath, amssymb, graphicx, physics, dsfont,array,subcaption}
\usepackage{orcidlink}
\usepackage{comment}
\usepackage{multirow}
\usepackage{float}
\usepackage{adjustbox}
\usepackage{float}

\usepackage{longtable} 
\usepackage{array}     

\usepackage[normalem]{ulem}

\allowdisplaybreaks

\usepackage{amsmath, bm}
\usepackage{graphicx}

\newcommand{\be}{\begin{equation}}
\newcommand{\ee}{\end{equation}}
\newcommand{\bes}{\begin{subequations} \begin{align} }
\newcommand{\ees}{\end{subequations}\end{align} }
\newcommand{\bea}{\begin{eqnarray}}
\newcommand{\eea}{\end{eqnarray}}

\newcommand{\fig}[1]{Fig.~\ref{fig:#1}}

\newcommand{\tab}[1]{Table~\ref{tab:#1}}

\def\bal#1\eal{\begin{align}#1\end{align}}

\newcounter{RSQ}

\newcounter{MSQ}

\title{\boldmath $\Upsilon(nS)$ Production within Jets at the LHC}

\author[a]{Taewook Ha\orcidlink{000-0002-5725-4007},}
\author[b]{Hee Sok Chung\orcidlink{0000-0003-1628-9315},}
\author[c]{Daekyoung Kang\orcidlink{0000-0002-9145-6913}}
\author[d]{Yunlu Wang\orcidlink{0009-0003-1313-1093},}
\author[a]{Haixiang Zhu\orcidlink{0009-0008-7084-7924},}

\affiliation[a]{Key Laboratory of Nuclear Physics and Ion-beam Application (MOE) and Institute of Modern Physics, Fudan University, Shanghai, China 200433}
\affiliation[b]{School of Mathematics and Physics, Kangwon National University, Gangneung 25457, Korea}
\affiliation[c]{
Department of Physics, Sejong University,Seoul 05006, Korea}
\affiliation[d]{College of Physics, Chengdu University of Technology, Chengdu 610059, China}

\emailAdd{taewookha@fudan.edu.cn}
\emailAdd{heesokchung@gwnu.ac.kr}
\emailAdd{dkang@sejong.ac.kr}
\emailAdd{ylwang516@cdut.edu.cn}
\emailAdd{25110200047@m.fudan.edu.cn}

\abstract{
Heavy quarkonium production inside jets offers a sensitive probe of QCD dynamics and bound-state formation mechanisms. While recent studies demonstrate that charmonium-in-jet observables effectively discriminate among competing nonrelativistic QCD (NRQCD) long-distance matrix element (LDME) sets, whether this discriminating power persists in the bottomonium sector remains an open question. Here, we present the first phenomenological study of $\Upsilon(1S)$, $\Upsilon(2S)$, and $\Upsilon(3S)$ production inside jets using the fragmenting jet function (FJF) framework at next-to-leading order (NLO), incorporating DGLAP evolution, threshold resummation, and feeddown contributions from higher bottomonium states. In sharp contrast to charmonium, we find that bottomonium-in-jet momentum-fraction ($z_H$) distributions exhibit a universal shape that is remarkably insensitive to the choice of LDME sets. We show that this universality stems from the strong dominance of the S-wave spin-triplet color-octet ($^3S_1^{[8]}$) production mechanism reinforced by $\chi_b$ feeddown transitions. Our predictions capture both the characteristic large-$z_H$ peak and the spectral broadening with increasing jet transverse momentum observed in recent CMS measurements. These results establish a clear physical distinction between charmonium and bottomonium fragmentation inside jets, providing a theoretical benchmark for future high-precision measurements at the LHC.}

\begin{document}
\maketitle

\section{Introduction}
\label{sec:1}

Measuring the quarkonium momentum fraction inside a jet~\cite{Baumgart:2014upa,Bain:2016clc}
yields information complementary to inclusive cross sections, providing deeper insight into the underlying production mechanisms.
Within NRQCD framework~\cite{Bodwin:1994jh}, the production of a physical quarkonium state is described as a sum of contributions from different intermediate heavy-quark pair configurations, each weighted by a corresponding LDME. Although NRQCD has achieved considerable phenomenological success, significant uncertainties remain regarding the relative importance of the various color-octet production channels. These ambiguities are fundamentally rooted in the inherent partial degeneracies of inclusive cross sections, which hinder the full disentanglement of distinct production mechanisms.
Consequently, such limitations are reflected in the substantial differences among existing LDME sets obtained from fits to inclusive production data~\cite{Brambilla:2022ayc,Brambilla:2021abf,Gong:2013qka,Feng:2015wka,Han:2014kxa}.

The momentum-fraction distribution of quarkonium inside jets has been recognized as a promising observable for studying the production mechanisms~\cite{Baumgart:2014upa}. Subsequent phenomenological studies demonstrated that different mechanisms generate characteristically different fragmentation patterns inside jets~\cite{Bain:2016clc,Bain:2017wvk,Wang:2025drz,Wang:2026dul,Copeland:2025osx}.  
Whether these distinct fragmentation patterns can be sufficiently resolved to discriminate among competing production scenarios, however, remains an open question, motivating further systematic investigation.

Corresponding measurements for charmonium by the LHCb~\cite{LHCb:2017llq,LHCb:2024ybz} and CMS~\cite{CMS:2019ebt,CMS:2021puf} Collaborations have provided direct experimental access to these distributions. In parallel, predictions based on \textsc{Pythia} with newly implemented quarkonium parton shower process \cite{Cooke:2023ldt} have also been presented for this observable \cite{LHCb:2024ybz,ValenciaPalomo:2025ljc,CMS:2026hnn}.

In addition to the LHC, typical jets at the EIC are predominantly quark-initiated~\cite{AbdulKhalek:2021gbh,Accardi:2012qut,Ee:2025scz,Zhu:2021xjn,Chu:2022jgs,Kang:2014qba,Kang:2013nha}. Hence, the quarkonium-in-jet observables offer a complementary probe of charmonium production~\cite{Wang:2026dul,EICph:2026hgp}.

The theoretical description of quarkonium production inside jets can be systematically formulated using the FJF framework~\cite{Procura:2009vm,Jain:2011xz,Baumgart:2014upa,Kang:2016ehg}, which combines perturbative jet dynamics with fragmentation functions (FFs) describing quarkonium formation. 
Beyond the longitudinal momentum fraction, the FJF formalism can be extended to 3D transverse-momentum-dependent observables, providing complementary probes of the production mechanisms~\cite{Kang:2017glf,Copeland:2023wbu}.

Within the FJF formalism, particular theoretical challenges arise when the observed quarkonium carries a large fraction of the jet momentum. In this kinematic regime, the available phase space for additional radiation becomes strongly restricted, generating large threshold logarithms in the short-distance coefficients (SDCs) of the quarkonium FFs. These logarithmically enhanced contributions can significantly modify the predicted distributions. A consistent treatment therefore requires resummation of these logarithms to obtain reliable theoretical predictions near the threshold region~\cite{Chung:2025gjk,Chung:2026mii}.

In previous studies of charmonium production inside jets~\cite{Wang:2025drz,Wang:2026dul}, we developed a framework incorporating threshold-resummed FFs within the FJF formalism and extended the accuracy of prediction beyond leading order in perturbation theory. We showed that our results can significantly improve the predicted momentum-fraction distributions and that quarkonium-in-jet observables can better discriminate among different LDME sets. %

Compared with the charmonium sector, bottomonium sector offers several theoretical advantages. The larger bottom-quark mass provides a harder perturbative scale for QCD calculations, and the substantially smaller heavy-quark velocity parameter improves the convergence of the NRQCD velocity expansion compared with charmonium~\cite{Bodwin:1994jh,Brambilla:2010cs}.

In the present work, we extend this framework to bottomonium production, perform a systematic study of $\Upsilon(nS)$ production inside jets, and present predictions for four representative LDME sets, with each set providing a distinct relative weights among the production mechanisms. To our knowledge, this represents the first phenomenological study of bottomonium production inside jets within the FJF framework incorporating threshold resummation at NLO.

Unlike the discriminating potential observed in charmonium production inside jets, we find that the distributions for bottomonium display a universal shape and are insensitive to the choice of LDME set, despite the sizable differences among the LDME sets. This universal shape is dominated by a single channel, $^3S^{[8]}_1$, while the contributions of the other production channels are relatively suppressed.

We compare our predictions with recent CMS measurements for $\Upsilon(1S)$, $\Upsilon(2S)$, and $\Upsilon(3S)$~\cite{CMS:2026hnn} and find that our predictions successfully describe the main features of the CMS measurement.
In addition, we present predictions for LHCb and explore different kinematic setups, such as momentum-fraction distributions at fixed quarkonium transverse momentum, as well as the dependence on the jet radius parameter.

The remainder of this paper is organized as follows. In Sec.~\ref{sec:2}, we summarize the theoretical framework, including the FJF factorization formalism and the implementation of threshold-resummed FFs. Section~\ref{sec:3} reviews the characteristic features of bottomonium production and introduces the NRQCD inputs used in the analysis, including the LDME classification and feeddown contributions. Numerical results are presented in Sec.~\ref{sec:4}, including comparisons with CMS data, studies of feeddown effects and NRQCD production channels, predictions for LHCb kinematics, and predictions for distributions at fixed quarkonium momentum and for the jet-radius dependence. Finally, Sec.~\ref{sec:5} contains our conclusions.

\section{Theoretical framework}
\label{sec:2}

In this section, we summarize the theoretical framework for describing $\Upsilon$ production inside jets. The calculation is based on the semi-inclusive FJF formalism~\cite{Kaufmann:2015hma, Kang:2016mcy, Kang:2016ehg} within Soft-Collinear Effective Theory (SCET)~\cite{Bauer:2000yr,Bauer:2001ct} combined with NRQCD FFs.

\subsection{Factorization formula}
\label{sec:2-1}

The production of a quarkonium state inside a jet involves several well-separated physical scales,
\[
\mu_H\sim p_T,\qquad
\mu_J\sim p_TR,\qquad
\mu_0\sim2m_b,
\]
corresponding to the hard-scattering scale, the jet scale, and the NRQCD matching scale, respectively. Here $p_T$ is the jet transverse momentum, $R$ is the jet radius, and $m_b$ is the bottom-quark mass. Their separation motivates the factorized description provided by the FJF formalism, which systematically separates hard scattering, jet dynamics, and quarkonium fragmentation.

Within the FJF framework, the differential
cross section for the inclusive process
$pp\to(\mathrm{jet}\,H)+X$ can be factorized as

\begin{equation}
\frac{d\sigma_{pp\to(\mathrm{jet}\,H)+X}}
     {dp_T\,d\eta\,dz_H}
=
\sum_i
\int_{z_{\min}}^1
\frac{dz}{z}\,
\frac{d\sigma^{i}}
     {dp_T^i\,d\eta}
(p_T^i,\eta,\mu)
\,
\mathcal{G}_i^H
(z,z_H,p_TR,\mu),
\label{eq:hadronic_sigma}
\end{equation}
where $p_T$ denotes the transverse momentum of the reconstructed jet and $p_T^i=p_T/z$ is the transverse momentum of the initiating parton $i$. The function $\mathcal{G}_i^H$ is the FJF, which describes the production of the quarkonium state $H$ inside a jet initiated by parton $i$.

The inclusive production cross section of the initiating parton $i$, including the convolution with the proton PDFs, is defined as

\begin{equation}
\frac{d\sigma^{i}}
     {dp_T^i\,d\eta}
(p_T^i,\eta,\mu)
=
\sum_{a,b}
\int dx_a\,dx_b\,
f_{a/p}(x_a,\mu)
f_{b/p}(x_b,\mu)
\,
\frac{d\hat{\sigma}_{ab\to i+X}}
     {dp_T^i\,d\eta}
(p_T^i,\eta,\mu),
\label{eq:parton_production}
\end{equation}
where $d\hat{\sigma}_{ab\to i+X}$ denotes the perturbatively calculable hard-scattering cross section for producing parton $i$, and $f_{a/p}$ and $f_{b/p}$ are the proton PDFs. A detailed derivation of the above factorized form, starting from the conventional factorization formula and performing the relevant kinematic transformations, is presented in Appendix~\ref{app:factorization}.

The variables $z$ and $z_H$ are the momentum fractions associated with jet production and quarkonium fragmentation, respectively, defined as

\begin{equation}
z = \frac{p_T}{p_T^i},
\qquad
z_H = \frac{p_T^H}{p_T}.
\label{eq:pt_fractions}
\end{equation}

Here $p_T^H$ denotes the transverse momentum of the observed quarkonium state. Throughout this work, we define the momentum fractions using these expressions.\footnote{In~\cite{Kaufmann:2015hma, Kang:2016mcy, Kang:2016ehg}, FJFs are  defined as light-cone energy fractions, and in the high-$p_T$ and narrow-jet limit, their difference is power suppressed.}

The FJF can be matched onto FFs as
\begin{equation}
\mathcal{G}_i^H(z,z_H,p_TR,\mu)
=
\sum_j
\int_{z_H}^{1}
\frac{d\xi}{\xi}
\,
\mathcal{J}_{ij}
\!\left(
z,\frac{z_H}{\xi},p_TR,\mu
\right)
D_j^H(\xi,\mu),
\label{eq:siFJF_matching}
\end{equation}
where $\mathcal{J}_{ij}$ are perturbatively calculable matching coefficients, while $D_j^H$ denotes the FF describing quarkonium production from parton $j$. The matching coefficients encode the perturbative dynamics associated with jet formation, whereas the FFs describe the nonperturbative formation of the observed quarkonium state.

Within NRQCD framework, the FFs can be factorized as
\begin{equation}
D_i^H(z_H,\mu_0)
=
\sum_n
d_{i\to Q\bar Q[n]}(z_H,\mu_0)
\,
\langle {\cal O}^H(n)\rangle ,
\label{eq:NRQCDFF}
\end{equation}
where $n$ labels the intermediate $Q\bar Q$ state, specified by its spectroscopic quantum numbers and color configuration, while $d_{i\to Q\bar Q[n]}$ are perturbatively calculable SDCs and $\langle {\cal O}^H(n)\rangle$ are the corresponding LDMEs. In this work, we retain only the leading color-singlet and color-octet channels, namely ${}^3S_1^{[1]}$, ${}^3S_1^{[8]}$, ${}^3P_J^{[8]}$, and ${}^1S_0^{[8]}$.

For the fixed-order component, we employ the SDCs computed through NLO in $\alpha_s$~\cite{Ma:2013yla}, building upon the earlier work of Refs.~\cite{Braaten:1994vv,Braaten:1996rp,Braaten:2000pc}. While NNLO corrections for the ${}^3S_1^{[8]}$ channel have recently become available~\cite{Feng:2026owq}, we consistently maintain NLO accuracy across all channels in this study.

\subsection{DGLAP evolution and threshold resummation}
\label{sec:2-2}

The scale dependence of the FFs is governed by the timelike DGLAP evolution equations~\cite{Gribov:1972ri,Altarelli:1977zs,Dokshitzer:1977sg},
\begin{equation}
\frac{d}{d\ln\mu^2}
D_i^H(z_H,\mu)
=
\sum_j
P_{ij}(z_H,\alpha_s)
\otimes
D_j^H(z_H,\mu),
\end{equation}
where $P_{ij}$ are the timelike splitting kernels and $\otimes$ denotes the standard convolution integral with respect to the momentum fraction. In this work, we employ the quark singlet and non-singlet decomposition of the evolution equations following Ref.~\cite{Vogt:2004ns}.

Starting from the NRQCD matching scale $\mu_0$, the FFs are evolved to the jet scale $\mu_J$, thereby resumming collinear logarithms associated with the ratio $\mu_J/\mu_0$. At the scale $\mu_J$, they are matched onto the FJFs through Eq.~(\ref{eq:siFJF_matching}). The resulting FJFs are subsequently evolved to the hard scale $\mu_H$ using timelike DGLAP evolution, resumming logarithms associated with the hierarchy $\mu_H \gg \mu_J \gg \mu_0$. The logarithms are resummed at leading-log (LL) accuracy in the present analysis. Beyond LL accuracy, it is known that higher-order terms in semi-inclusive jet functions modify the evolution~\cite{Lee:2024icn,Lee:2024tzc}.

In addition to collinear logarithms resummed through DGLAP evolution, large threshold logarithms arise in the limit $z_H \to 1$, where soft-gluon radiation is kinematically suppressed. To account for these effects, we employ threshold-resummed fragmentation-function SDCs at the matching scale $\mu_0$, following the formalism developed in Refs.~\cite{Chung:2024jfk,Chung:2026mii} and applied in Refs.~\cite{Wang:2025drz,Wang:2026dul}. The resummation is applied to the gluon-induced contributions to the ${}^3S_1^{[8]}$ and ${}^3P_J^{[8]}$ channels, as well as to the ${}^3P_J^{[1]}$ channels relevant for $\chi_b$ feeddown to the $\Upsilon$.
The resummation is performed in Mellin space, where threshold logarithms exponentiate and convolution integrals reduce to products. The SDCs are supplemented with leading threshold double logarithms, which improve the convergence in the large-$z_H$ region.

For completeness, a comparison between the threshold-resummed and fixed-order predictions is presented in Appendix~\ref{app:FOcomparison}. As shown there, the resummation has only a modest numerical impact over most of the $z_H$ range and becomes significant primarily in the threshold region, where large logarithmic corrections are expected.

\section{Bottomonium matrix elements}
\label{sec:3}

This section discusses the nonperturbative NRQCD LDME sets employed in the paper and feeddown contributions from excited bottomonium states.

\subsection{LDME classification}
\label{sec:3-1}

In the NRQCD framework, the relative importance of different production channels is governed by the heavy-quark velocity expansion~\cite{Bodwin:1994jh}. The characteristic velocity parameter is approximately

\[
v_c^2\sim0.3,
\qquad
v_b^2\sim0.1,
\]
indicating a substantially more nonrelativistic system in bottomonium than in charmonium.

The NRQCD velocity expansion is realized through a hierarchy of four-fermion operators, each associated with a distinct production channel. The corresponding LDMEs encode the nonperturbative transition of the heavy-quark pair into the observed quarkonium state. For S-wave quarkonium production, the velocity-scaling rules imply
\[
\langle O(^3S_1^{[1]})\rangle
\sim
m_Q^3v^3,
\]
whereas the leading color-octet matrix elements satisfy
\[
\langle O(^3S_1^{[8]})\rangle,
~
\langle O(^1S_0^{[8]})\rangle,
~
\langle O(^3P_J^{[8]})\rangle/m_Q^2
\sim
m_Q^3v^7.
\]

\begin{table*}[t]
\centering
\footnotesize
\setlength{\tabcolsep}{4pt}
\renewcommand{\arraystretch}{1.15}

\begin{tabular}{|c|c|c|c|c|c|c|}
\hline
\multirow{2}{*}{\textbf{Category}}
&
\multirow{2}{*}{\textbf{State}}
&
\multirow{2}{*}{\textbf{Reference}}
&
$\langle O(^3S_1^{[1]})\rangle$
&
$\langle O(^3S_1^{[8]})\rangle$
&
$\langle O(^1S_0^{[8]})\rangle$
&
$\langle O(^3P_0^{[8]})\rangle/m_Q^2$
\\

&
&
&
(GeV$^3$)
&
($10^{-2}$ GeV$^3$)
&
($10^{-2}$ GeV$^3$)
&
($10^{-2}$ GeV$^3$)
\\
\hline

\multirow{2}{*}{1}
& $\Upsilon(1S)$
& Brambilla \textit{et al.}~\cite{Brambilla:2022ayc}
& $9.28\pm0.93$
& $2.96\pm0.93$
& $-0.40\pm2.04$
& $2.12\pm0.68$
\\

& $J/\psi$
& Brambilla \textit{et al.}~\cite{Brambilla:2022ayc}
& $1.18\pm0.35$
& $1.40\pm0.42$
& $-0.63\pm3.22$
& $2.33\pm0.83$
\\
\hline

\multirow{2}{*}{2}
& $\Upsilon(1S)$
& Gong \textit{et al.}~\cite{Gong:2013qka}
& $9.28$
& $-0.76\pm0.24$
& $11.15\pm0.43$
& $-0.67$
\\

& $J/\psi$
& Bodwin \textit{et al.}~\cite{Bodwin:2015iua}
& $1.32\pm0.20$
& $-0.71\pm0.36$
& $11.0\pm1.4$
& $-0.31\pm0.15$
\\
\hline

\multirow{2}{*}{3}
& $\Upsilon(1S)$
& Feng \textit{et al.}~\cite{Feng:2015wka}
& $9.28$
& $0.47\pm0.41$
& $11.6\pm2.61$
& $-0.49\pm0.59$
\\

& $J/\psi$
& B\&K~\cite{Butenschoen:2011yh}
& $1.32\pm0.20$
& $0.22\pm0.06$
& $4.97\pm0.44$
& $-0.72\pm0.09$
\\
\hline

4
& $\Upsilon(1S)$
& Han \textit{et al.}~\cite{Han:2014kxa}
& $9.28$
& $1.17\pm0.02$
& $13.7\pm1.1$
& $0$
\\
\hline

\end{tabular}

\caption{
Representative NRQCD LDME sets for direct $\Upsilon(1S)$ and $J/\psi$ production.
All quarkonium LDMEs are quoted at the scale $\mu_\Lambda=m_Q$.}
\label{tab:1SwaveLDMEs}

\end{table*}

This qualitative expectation is reflected in Table~\ref{tab:1SwaveLDMEs}, which compares representative LDME sets for direct $J/\psi$ and $\Upsilon(1S)$ production. 
The LDMEs for the excited $\Upsilon(nS)$ states are listed in Appendix~\ref{app:LDME}.
The overall hierarchy between the color-singlet and color-octet channels is broadly consistent with the qualitative expectation from the NRQCD velocity-scaling rules.

Despite the hierarchy implied by the velocity-scaling rules, different global analyses still yield substantially different combinations of color-octet matrix elements, leading to noticeably different predictions for quarkonium production inside jets. Rather than discussing each fit individually, it is more useful to group the available LDME sets according to their relative sign and strength. This classification provides a convenient framework for interpreting our numerical results, which span as many cases in the literature as possible.

Since the SDC of the ${}^3P_J^{[8]}$ channel is negative over most of the relevant $z_H$ range in our calculation, the relative signs of $\langle O(^3S_1^{[8]})\rangle$ and $\langle O(^3P_J^{[8]})\rangle$ provide the primary criterion for distinguishing different LDME sets, determining whether these two channels contribute constructively or destructively. Once the interference pattern is fixed, the size of the ${}^1S_0^{[8]}$ matrix element further refines the classification. Based on these considerations, we can classify the $\Upsilon$ LDME sets proposed in the literature~\cite{Gong:2013qka,Han:2014kxa,Feng:2015wka,Brambilla:2022ayc,Abdulov:2020nxh,Braaten:2000cm,Cho:1995ce,Sharma:2012dy,Abdulov:2019uyx,Abdulov:2020edt,Sun:2012vc,Schuler:1997is} into four distinct categories.

\begin{itemize}

\item \textbf{Category 1:}
$\langle O(^3S_1^{[8]})\rangle$
and
$\langle O(^3P_J^{[8]})\rangle$
have the same sign, while
$\langle O(^1S_0^{[8]})\rangle$
is relatively small.

\item \textbf{Category 2:}
$\langle O(^3S_1^{[8]})\rangle$
and
$\langle O(^3P_J^{[8]})\rangle$
have the same sign, while
$\langle O(^1S_0^{[8]})\rangle$
is relatively large.

\item \textbf{Category 3:}
$\langle O(^3S_1^{[8]})\rangle$
and
$\langle O(^3P_J^{[8]})\rangle$
have opposite signs.

\item \textbf{Category 4:}
$\langle O(^3P_J^{[8]})\rangle$
is minimized and $\langle O(^1S_0^{[8]})\rangle$ is maximized.
\end{itemize}

For $\Upsilon(1S)$ production, we employ representative LDME sets from all four categories~\cite{Gong:2013qka,Han:2014kxa,Feng:2015wka,Brambilla:2022ayc}: Brambilla (Category~1), Gong (Category~2), Feng (Category~3), and Han (Category~4) as shown in \tab{1SwaveLDMEs}.

The Brambilla set is based on the approximate universality
relation of LDMEs among $S$-wave quarkonia, which is used as a constraint in
a combined fit on charmonium and bottomonium production data.
This universality relation, which was first discovered
in Refs.~\cite{Brambilla:2022rjd, Brambilla:2022ayc} in the potential NRQCD
(pNRQCD) formalism and recently rederived in Ref.~\cite{Copeland:2026yqa} using
the combined velocity NRQCD (vNRQCD) and pNRQCD approach, implies that the
color-octet LDMEs are factorized into the quarkonium wavefunction at the
origin, multiplied by vacuum expectation values of gluonic operators, the
latter of which are independent of the heavy quark flavor and radial excitation
of the quarkonium state. In this case, the relative sizes of the color-octet
LDMEs compared to $\langle O(^3S_1^{[1]})\rangle$ are independent of the radial
excitation. Note that, due to the approximate degeneracy in the $p_T$ shape of
the quarkonium production SDCs as pointed out in Ref.~\cite{Han:2014kxa}, the
universality relation does not arise automatically from fits to individual
quarkonium data; because of this, most LDME sets in the literature do not
satisfy this universality relation.

The category~2 is represented by Gong LDME set, in which both $\langle O(^3S_1^{[8]})\rangle$ and $\langle O(^3P_J^{[8]})\rangle$ are negative. The Gong LDME set was originally extracted at the NRQCD scale $\mu_\Lambda=m_bv$. For consistency with the other bottomonium LDME sets, we evolve it to the common scale $\mu_\Lambda=m_b$ using the fixed-order NRQCD renormalization-group equation before performing the phenomenological analysis. Further details on this RG solution are provided in Appendix~\ref{app:LDME}.

For the Han set, we adopt the representative solution obtained by maximizing $\langle O(^1S_0^{[8]})\rangle$ under the positivity constraint. In this representative solution, $\langle O(^3P_J^{[8]})\rangle$ takes its minimum value, namely zero, and is therefore classified as Category~4.

Since the NRQCD matrix elements are fitted independently for each $\Upsilon(nS)$ state, the corresponding LDME sets do not necessarily belong to the same category for different states. In particular, among the currently available determinations, Category~2 is realized only for the $\Upsilon(1S)$ state, and no representative Category~2 set is available for $\Upsilon(2S)$ or $\Upsilon(3S)$.

The complete sets of LDMEs adopted in this work, including those for $\Upsilon(1S)$, $\Upsilon(2S)$, and $\Upsilon(3S)$ states, are listed in Appendix~\ref{app:LDME}.

\subsection{Feeddown contribution}
\label{sec:3-2}

The inclusive production includes both direct production and feeddown contributions from higher bottomonium states. A characteristic feature of bottomonium production is the presence of several low-lying excited states that undergo radiative or hadronic transitions to lower $\Upsilon(nS)$ states. Such transitions constitute an essential component of inclusive $\Upsilon$ production and must therefore be incorporated for a quantitative comparison with experimental measurements.

\begin{table}[t]
\centering
\small
\renewcommand{\arraystretch}{1.15}
\begin{adjustbox}{center}
\begin{tabular}{c|ccccccccc}
\hline
Parent state $H$
& $\Upsilon(2S)$
& $\Upsilon(3S)$
& $\chi_{b1}(1P)$
& $\chi_{b2}(1P)$
& $\chi_{b1}(2P)$
& $\chi_{b2}(2P)$
& $\chi_{b1}(3P)$
& $\chi_{b2}(3P)$
\\
\hline
$\mathrm{Br}(H\rightarrow\Upsilon(1S))$
& 0.265
& -
& 0.352
& 0.180
& 0.115
& 0.077
& 0.038
& -
\\

$\mathrm{Br}(H\rightarrow\Upsilon(2S))$
& -
& 0.106
& -
& -
& 0.181
& 0.089
& 0.037
& -
\\

$\mathrm{Br}(H\rightarrow\Upsilon(3S))$
& -
& -
& -
& -
& -
& -
& 0.104
& 0.061
\\
\hline
\end{tabular}
\end{adjustbox}
\caption{
Branching fractions for feeddown transitions included in the inclusive $\Upsilon(nS)$ calculations. The $\chi_{b0}$ transitions are omitted due to their negligible contributions.
}
\label{tab:feeddown}
\end{table}

The representative feeddown channels included in the present calculation are summarized in Table~\ref{tab:feeddown}. Branching fractions for the established states are taken from the Particle Data Group~\cite{ParticleDataGroup:2024cfk}, while those for the $\chi_b(3P)$ transitions are taken from the predictions of Ref.~\cite{Han:2014kxa}. The $\chi_b(nP)$ LDMEs adopted in this work are listed in Appendix~\ref{app:LDME}. Since the $\chi_b(3P)$ states were not included in the LDME extraction of Ref.~\cite{Gong:2013qka}, the corresponding feeddown contribution is omitted when using the Gong LDME set.

Unless otherwise stated, the parent bottomonium states are taken to be inclusive, so that cascade transitions such as $\chi_b(2P)\rightarrow\Upsilon(2S)\rightarrow\Upsilon(1S)$ are consistently included. The feeddown channels included in our calculation are selected according to their effective contribution to the inclusive yield. For $\chi_b$ feeddown channels, we retain states satisfying $Br(\chi_{bJ}\rightarrow\Upsilon+X)(2J+1)>0.1$, where the factor $(2J+1)$ accounts for the spin degeneracy of the $\chi_{bJ}$ multiplet. For $nS\rightarrow n'S$ transitions, the corresponding branching fractions alone are used to determine their relevance. Channels below these criteria are not included in the calculation.

Beyond its contribution to the production rate, feeddown also modifies the shape of the $z_H$ distribution. This follows from the fact that the daughter quarkonium carries only a fraction of the parent's momentum, so that its momentum fraction with respect to the jet is correspondingly reduced. In the collinear limit,
\begin{equation}
\label{zH}
z_H \simeq \frac{m_H}{m_{H'}} z_{H'},
\end{equation}
where $H'$ denotes the parent bottomonium state and $H$ the observed daughter state. Because $m_{H'}>m_H$, feeddown shifts events toward smaller values of $z_H$, leading to a broadening of the fragmentation spectrum. This effect is most pronounced for $\Upsilon(1S)$, while it is reduced for both $\Upsilon(2S)$ and $\Upsilon(3S)$.

\section{Numerical results}
\label{sec:4}

We present numerical results for $\Upsilon$ production inside jets, first focusing on the structure of direct production and its dependence on the choice of LDME set, and then examining the impact of feeddown to compare the resulting inclusive $\Upsilon$ predictions with the CMS measurements for $\Upsilon(1S)$, $\Upsilon(2S)$, and $\Upsilon(3S)$ production.

Unless otherwise stated, the numerical results presented below are obtained using the following setup. The perturbative hard-scattering cross sections are computed within the INCNLO framework~\cite{Aurenche:1999nz}. The perturbative scale is chosen as $\mu=p_T$, while the jet scale is taken as $\mu_J=p_TR$. The theoretical predictions are evaluated within the same fiducial phase space as the CMS measurement~\cite{CMS:2026hnn}, corresponding to proton-proton collisions at $\sqrt{s}=13$ TeV with jets reconstructed using the anti-$k_T$ algorithm with radius parameter $R=0.4$ and restricted to the central region $|\eta|<1.5$.

\subsection{Direct production}
\label{sec:4-1}

We focus on direct $\Upsilon(1S)$ production as a benchmark case because it exhibits the largest sensitivity to the choice of LDME set. Although the direct $\Upsilon(2S)$ and $\Upsilon(3S)$ spectra show qualitatively similar channel patterns, the interplay among the color-octet contributions is most clearly demonstrated in the $\Upsilon(1S)$ case.

\begin{figure}[tb]
    \centering
    \includegraphics[width=0.8\textwidth]{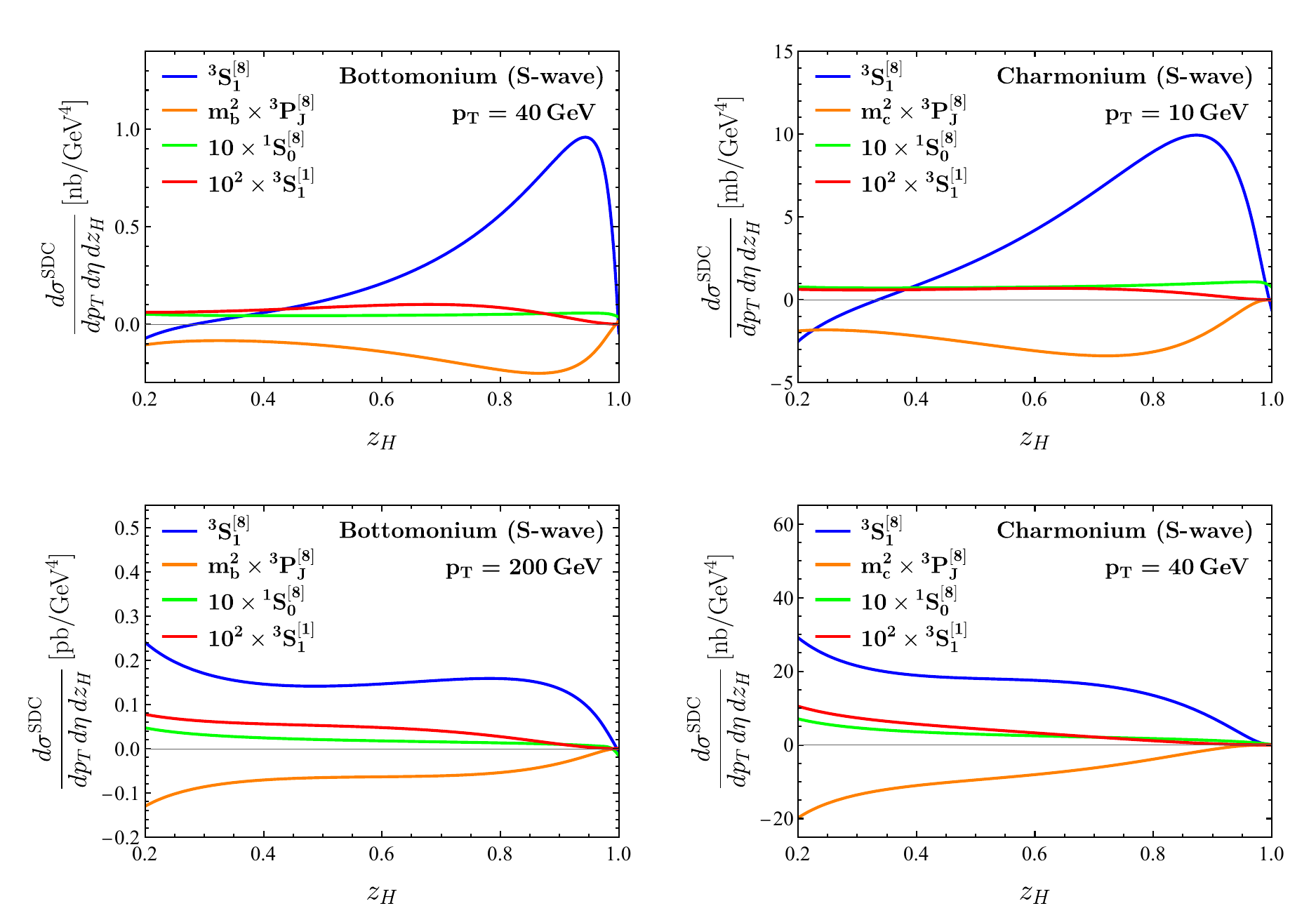}
    \vspace{-2ex}
\caption{
SDCs for individual channels as a function of $z_H$ at different transverse momentum scales. The left and right columns correspond to bottomonium and charmonium S-wave production, respectively. The $^{1}S_0^{[8]}$ ($^{3}S_1^{[1]}$) channels are scaled by factors of 10 ($10^2$) for visibility. The $^{3}P_J^{[8]}$ SDCs are multiplied by $m_Q^2$, where $m_c=1.5$ GeV and $m_b=4.75$ GeV.
}
    \label{fig:sdc_evolution}
\end{figure}

To examine the scale dependence of the channel structure, we decompose the SDCs into individual production channels as a function of $p_T$, as shown in \fig{sdc_evolution} for charmonium and bottomonium production. For the ${}^3P_J^{[8]}$ channel, we display $m_Q^2$-rescaled SDCs so that all channels have the same mass dimension.

At relatively low transverse momenta ($p_T=10$ GeV for charmonium and $p_T=40$ GeV for bottomonium), the ${}^3S_1^{[8]}$ channel dominates the threshold region due to the $\delta(1-z_H)$ contribution from the gluon fragmentation process. As $p_T$ increases, DGLAP evolution smears this threshold contribution toward smaller $z_H$ through gluon radiation, reducing the peak near $z_H\sim1$ and enhancing the low-$z_H$ region.
This evolution exhibits a pronounced dependence on the heavy-quark mass. For charmonium, the threshold enhancement is largely washed out already at moderate $p_T$, whereas for bottomonium the corresponding structure remains visible over the CMS kinematic range considered in this work. This difference reflects the reduced impact of DGLAP evolution for bottomonium due to its larger heavy-quark mass.

\begin{figure}[tb]
    \centering
    \begin{adjustbox}{center}
    \includegraphics[width=1.1\textwidth]{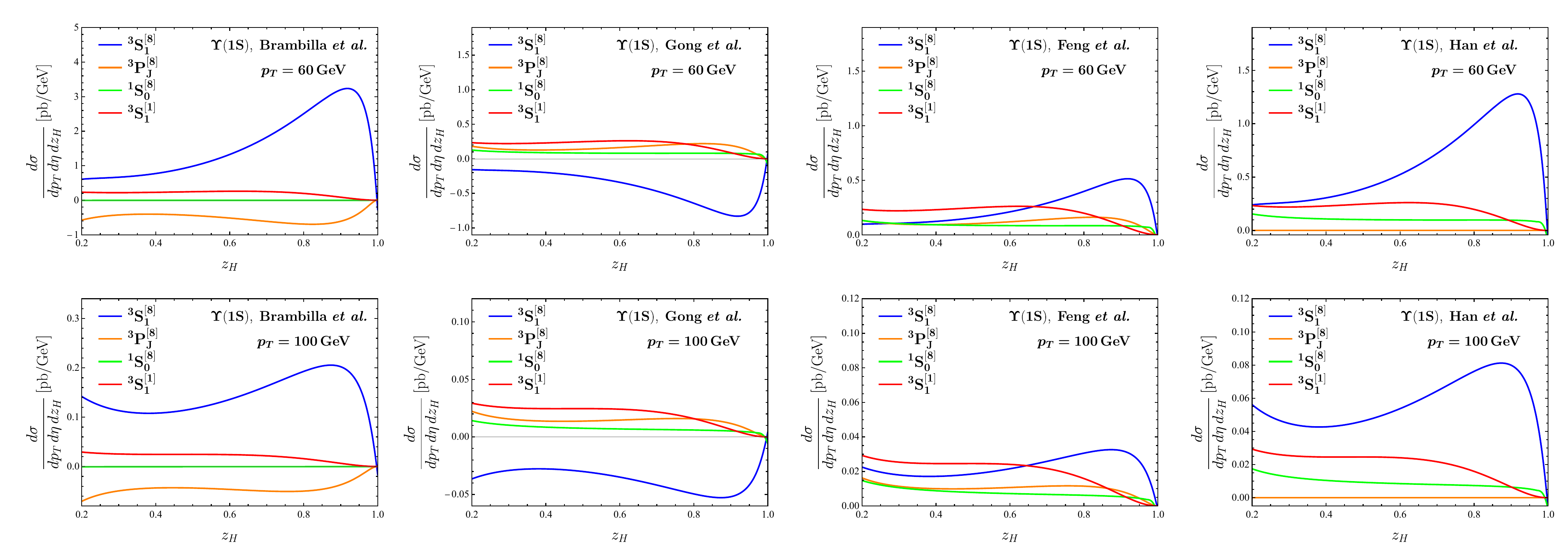}
    \end{adjustbox}
    \vspace{-2ex}
    \caption{The product of LDMEs and SDCs for direct $\Upsilon(1S)$ production.
    }
    \label{fig:LDMExSDC}
\end{figure}

To understand how the different LDME sets modify the relative contributions of the individual channels, we examine the products of SDCs and LDMEs shown in \fig{LDMExSDC}. The figure displays two different $p_T$ bins (rows) and four different LDME sets (columns). As discussed in Sec.~\ref{sec:3-1}, the ${}^3P_J^{[8]}$ channel can contribute constructively or destructively depending on the sign of its LDME. \fig{LDMExSDC} directly illustrates how the different LDME sets alter the interference pattern and the relative contributions of the channels.

The Brambilla LDME set exhibits a pronounced cancellation between the ${}^3S_1^{[8]}$ and ${}^3P_J^{[8]}$ channels.
The Gong LDME set exhibits a different cancellation pattern. It contains a negative $^{3}S_1^{[8]}$ LDME whose magnitude is substantially smaller than in the other sets. Consequently, the $^{1}S_0^{[8]}$ and $^{3}P_J^{[8]}$ channels play a relatively larger role in the total octet contribution. The negative $^{3}S_1^{[8]}$ contribution give rise to a negative direct-production contribution near $z_H=1$.
The Feng set exhibits constructive interference between the ${}^3S_1^{[8]}$ and ${}^3P_J^{[8]}$ channels, leading to a larger octet contribution.
The Han set is dominated by the ${}^3S_1^{[8]}$ channel, with no ${}^3P_J^{[8]}$ contribution.

For the Brambilla LDME set, the relative importance of the ${}^3S_1^{[8]}$ channel in bottomonium production can be understood from the NRQCD scale evolution of the color-octet matrix elements. As shown in Appendix~\ref{app:LDME}, the renormalization-group evolution induces a mixing between the ${}^3S_1^{[8]}$ LDME and the ${}^3P_J^{[8]}$ LDME. This scale evolution modifies the relative hierarchy among the color-octet contributions and contributes to the different channel structure observed in bottomonium compared with charmonium. In particular, while large cancellations between the ${}^3S_1^{[8]}$ and ${}^3P_J^{[8]}$ channels can occur for charmonium production~\cite{Brambilla:2022ayc}, the ${}^3S_1^{[8]}$ contribution remains relatively more prominent in the bottomonium case. A similar mixing occurs for $\chi_b$ production, where the ${}^3S_1^{[8]}$ LDME evolves through mixing with the ${}^3P_J^{[1]}$ LDME.

\subsection{Feeddown effects}
\label{sec:4-2}

\begin{figure}[tb]
\centering
\includegraphics[width=1.0\textwidth]{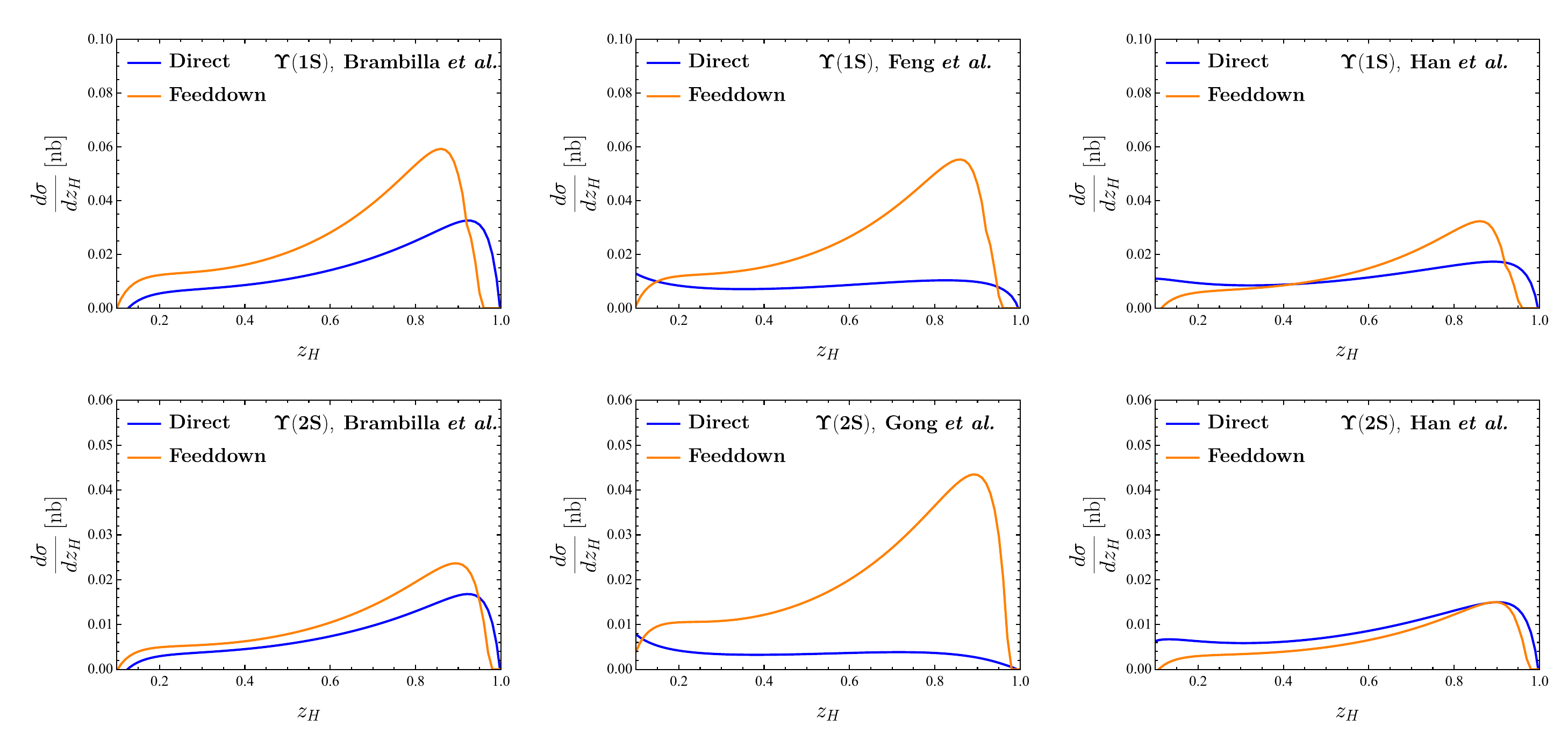}
\caption{
Comparison of the direct-production and feeddown contributions to the inclusive $\Upsilon(1S)$ (top) and $\Upsilon(2S)$ (bottom) $z_H$ spectra for $60<p_T<75$ GeV.
}
\label{fig:feeddownComparison}
\end{figure}

Feeddown contributions can substantially modify the composition of the inclusive production. \fig{feeddownComparison} compares the direct-production and feeddown contributions to the inclusive $\Upsilon(1S)$ and $\Upsilon(2S)$ spectra for the representative LDME sets.

An important feature of \fig{feeddownComparison} is that the feeddown contribution exhibits a universal shape among the different LDME sets, despite the substantial variations observed in the direct-production spectra. For both inclusive $\Upsilon(1S)$ and $\Upsilon(2S)$ production, feeddown is largely supplied by $\chi_b$ states.

In the kinematic region considered here, where the large-$z_H$ fragmentation contribution remains sizable, $\chi_b$ production receives a dominant contribution from the ${}^3S_1^{[8]}$ channel. This behavior can be understood from the relative contributions of the ${}^3S_1^{[8]}$ and ${}^3P_J^{[1]}$ channels. Although the corresponding matrix elements follow the same spin-degeneracy scaling, the ${}^3P_J^{[1]}$ contribution is found to be strongly suppressed after summing over the $\chi_{bJ}$ multiplet. As a result, the feeddown spectrum is dominated by the ${}^3S_1^{[8]}$ fragmentation channel.

The shift of the feeddown contribution toward smaller values of $z_H$ and its stronger suppression near $z_H=1$ can be traced back to the feeddown kinematics discussed in Sec.~\ref{sec:3-2}. In particular, Eq.~(\ref{zH}) shows that the daughter quarkonium inherits only a fraction of the parent momentum, resulting in a softer $z_H$ distribution.

The relative importance of feeddown compared with direct production depends on the Upsilon state. For $\Upsilon(1S)$, the feeddown contribution provides a substantial component of the total yield and becomes comparable to, or even larger than, the direct contribution over a significant part of the $z_H$ range. For $\Upsilon(2S)$, the feeddown contribution remains sizable, while the relative importance of the direct component depends on the chosen LDME set.
The corresponding comparison for $\Upsilon(3S)$, not shown separately, is very similar to that observed for $\Upsilon(2S)$.
Although the feeddown channels for $\Upsilon(3S)$ are restricted to the $\chi_b(3P)$ states, the sizable branching fractions of the $\chi_b(3P)\rightarrow\Upsilon(3S)$ transitions result in a non-negligible feeddown contribution.

\subsection{Comparison with CMS results}
\label{sec:4-3}

In order to compare with the CMS measurement, we apply the CMS fiducial muon selection at particle level by explicitly simulating the dimuon decay $\Upsilon\rightarrow\mu^+\mu^-$ for our prediction. The selection criteria applied at the muon and dimuon levels are:
$p_T^\mu > 6~{\rm GeV}$, $|\eta^\mu| < 1.4$, $p_T^{\mu\mu} > 15~{\rm GeV}$, and $|y^{\mu\mu}| < 1.2$.

\begin{figure*}[b]
\centering
\begin{adjustbox}{center}
\includegraphics[width=1.1\textwidth]{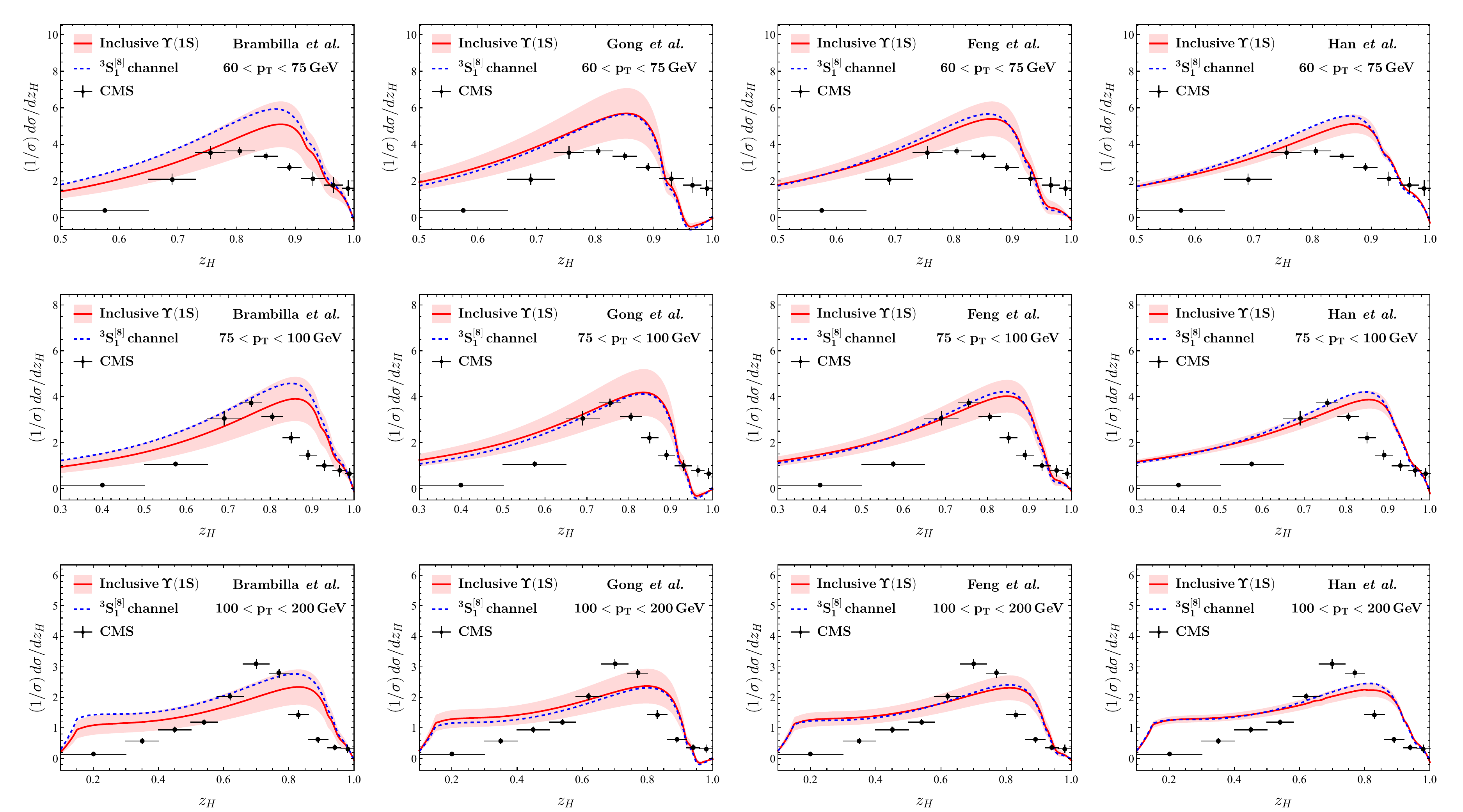}
\end{adjustbox}
\caption{
Comparison between the theoretical predictions and the CMS data for the normalized $z_H$ distribution of $\Upsilon(1S)$ production inside jets. The three rows correspond to $60 <p_T <75$ GeV, $75 < p_T < 100$ GeV, and $100 < p_T < 200$ GeV.}
\label{fig:CMScomparison1S}
\end{figure*}

These requirements significantly suppress the low-$z_H$ region, where the quarkonium carries only a small fraction of the jet transverse momentum. Since the reconstructed dimuon system originates from the $\Upsilon\rightarrow\mu^+\mu^-$ decay, its transverse momentum is equal to that of the parent quarkonium, $p_T^{\mu\mu}=p_T^H$. Therefore, the dimuon requirement effectively imposes a $p_T$-dependent lower bound on $z_H$. For example, for
$p_T=60~{\rm GeV}$,
the requirement
$p_T^{\mu\mu}>15~{\rm GeV}$
corresponds to
\[
z_H>\frac{15}{60}=0.25.
\]
In practice, the suppression extends over a broader region due to the finite $p_T$ bin and additional fiducial requirements. The resulting acceptance-induced suppression in the $z_H$ distribution is illustrated in \fig{mucut}. Its dependence on the jet transverse momentum is further discussed in Appendix.~\ref{app:mucut}.

\fig{CMScomparison1S} compares our theoretical predictions with the CMS data~\cite{CMS:2026hnn} for the normalized $z_H$ distribution. Both the inclusive production prediction (solid) and the $^3S_1^{[8]}$ channel contribution (dotted), incorporating direct and feeddown processes, are displayed.

The uncertainty bands shown in \fig{CMScomparison1S} include only the uncertainties associated with the NRQCD LDMEs. The individual color-octet LDME uncertainties are treated as independent variations and combined in quadrature. Perturbative scale variations are not included in this analysis, as our primary focus is to isolate the sensitivity of the normalized $z_H$ distributions to the choice of LDME set. For the Brambilla LDME set, we additionally checked the impact of using the covariance matrix. This alternative treatment yields somewhat narrower uncertainty bands, particularly in the low-$z_H$ region, but does not modify the qualitative behavior of the prediction or the conclusions of the comparison.

For the highest transverse-momentum bin, the CMS measurement is reported for $p_T > 100$ GeV, whereas our numerical calculation is performed over the finite range $100 < p_T < 200$ GeV. The contribution from higher transverse momenta is expected to be small and is neglected in the comparison. To facilitate a comparison of the distribution shapes, the normalization is performed over the region $z_H>z_{\rm cut}$, excluding the first two experimental bins. This prescription reduces the sensitivity of the normalization to the low-$z_H$ region, where additional experimental selection and kinematic effects can have a larger impact, and allows for a consistent comparison across all transverse-momentum bins.

Overall, the calculation reproduces the characteristic large-$z_H$ peak structure reasonably well and captures the qualitative $p_T$ dependence observed by CMS: as jet $p_T$ increases, the distributions shift toward smaller $z_H$ due to stronger fragmentation evolution.

The large-$z_H$ enhancement is driven by the ${}^{3}S_1^{[8]}$ fragmentation contribution, which reflects the threshold-enhanced structure of its gluon FF. In the inclusive spectrum, this feature is further reinforced by feeddown—primarily from $\chi_b$ states dominated by the $^{3}S_1^{[8]}$ channel. While direct production receives comparable contributions from different channels in Gong and Feng LDME sets (\fig{LDMExSDC}), feeddown is substantial and often exceeds the direct component across a wide $z_H$ range (\fig{feeddownComparison}). Consequently, the inclusive spectrum largely inherits the characteristic shape of the ${}^{3}S_1^{[8]}$ mechanism.

The subtle variations in peak position among the LDME sets arise from different relative weights of subleading channels. In particular, the interplay between the dominant $^{3}S_1^{[8]}$ and subleading $^{3}P_J^{[8]}$ channels alters the spectral shape around large $z_H$. For the Brambilla set, the negative $^{3}P_J^{[8]}$ contribution partially cancels the $^{3}S_1^{[8]}$ term, yielding a slightly harder distribution. For other sets, different balances among subleading channels induce minor shifts near the peak. Despite these channel-level variations, all LDME sets yield remarkably similar normalized spectra because the ${}^{3}S_1^{[8]}$ fragmentation mechanism remains universally dominant, with feeddown further stabilizing the distribution.

However, a noticeable discrepancy remains in the low-$z_H$ region after the fiducial acceptance corrections are included. The calculation also reproduces the qualitative $p_T$ dependence of the spectra, with the distributions extending toward smaller $z_H$ at higher jet transverse momenta due to the fragmentation evolution. However, the low-$z_H$ enhancement is systematically larger than observed in the CMS data.

The origin of this discrepancy is not fully understood within the current framework. One possible source is the treatment of heavy-quark mass effects, since the FJF used in this work is derived in the massless limit and finite-mass corrections are not included. In addition, missing higher-order perturbative corrections may affect the fragmentation evolution and the resulting $z_H$ dependence, although contributions from other perturbative ingredients cannot be excluded. A more complete treatment including finite-mass effects and higher-order corrections in the relevant perturbative ingredients may therefore be required for a more quantitative description of the measured spectrum.

\begin{figure}[tb]
\centering
\includegraphics[width=1\textwidth]{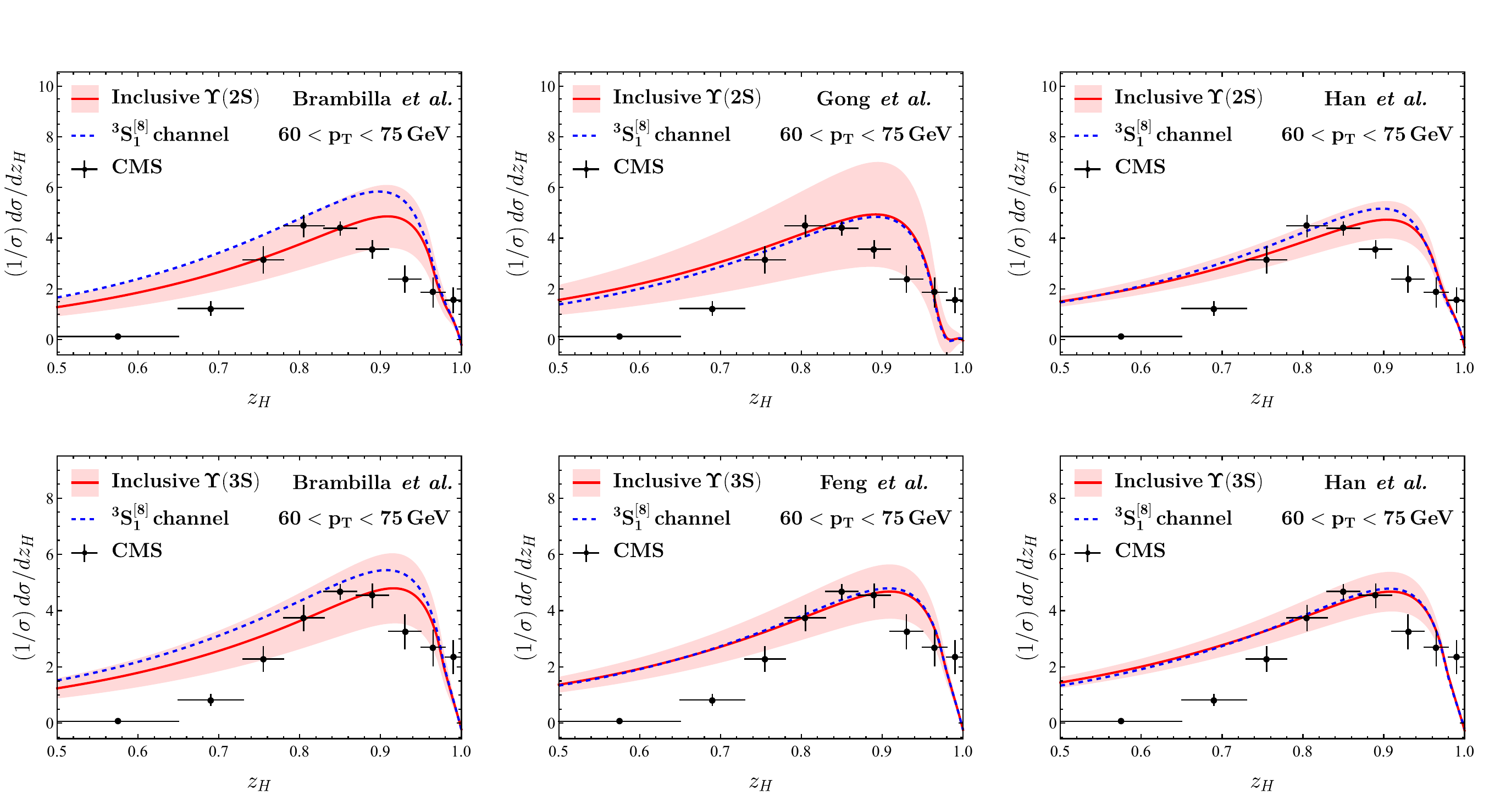}
\caption{
Comparison between the theoretical predictions and the CMS data for the normalized $z_H$ distributions of inclusive $\Upsilon(2S)$ (upper row) and $\Upsilon(3S)$ (lower row) production for different LDME sets. 
}
\label{fig:CMScomparison2S3S}
\end{figure}

%

\fig{CMScomparison2S3S} compares the predictions for inclusive $\Upsilon(2S)$ and $\Upsilon(3S)$ production with the CMS data using different LDME sets. Representative results for the Brambilla, Gong/Feng, and Han sets are shown, while comparisons for higher-$p_T$ bins, which exhibit similar qualitative behavior, are provided in Appendix~\ref{app:LDMEcomparison}.

The predictions reproduce the overall shape of the CMS data reasonably well in the moderate- and large-$z_H$ regions. A discrepancy remains in the low-$z_H$ region, with a behavior qualitatively similar to that observed for $\Upsilon(1S)$ production.

The distributions are remarkably similar among the different LDME sets, following the same trend observed for $\Upsilon(1S)$ production. This behavior reflects the dominant role of the ${}^{3}S_1^{[8]}$ fragmentation contribution in the excited-bottomonium spectra.

The approximate universality relations for the Brambilla LDME set discussed in Sec.~\ref{sec:3-1} provide a possible interpretation of the similar $z_H$ shapes among the different $\Upsilon(nS)$ states. The relations imply that the relative sizes of the color-octet LDMEs with respect to the color-singlet matrix element are approximately independent of the radial excitation, while the differences among the inclusive $\Upsilon(nS)$ distributions arise mainly from their different feeddown contributions. Together with the dominance of the ${}^3S_1^{[8]}$ channel in the relevant $\chi_b$ feeddown contributions, this provides a natural explanation for the similar shapes observed among the $\Upsilon(nS)$ states.

\subsection{Prediction for LHCb}
\label{sec:4-4}

\begin{figure}[tb]
\centering
\includegraphics[width=0.8\textwidth]{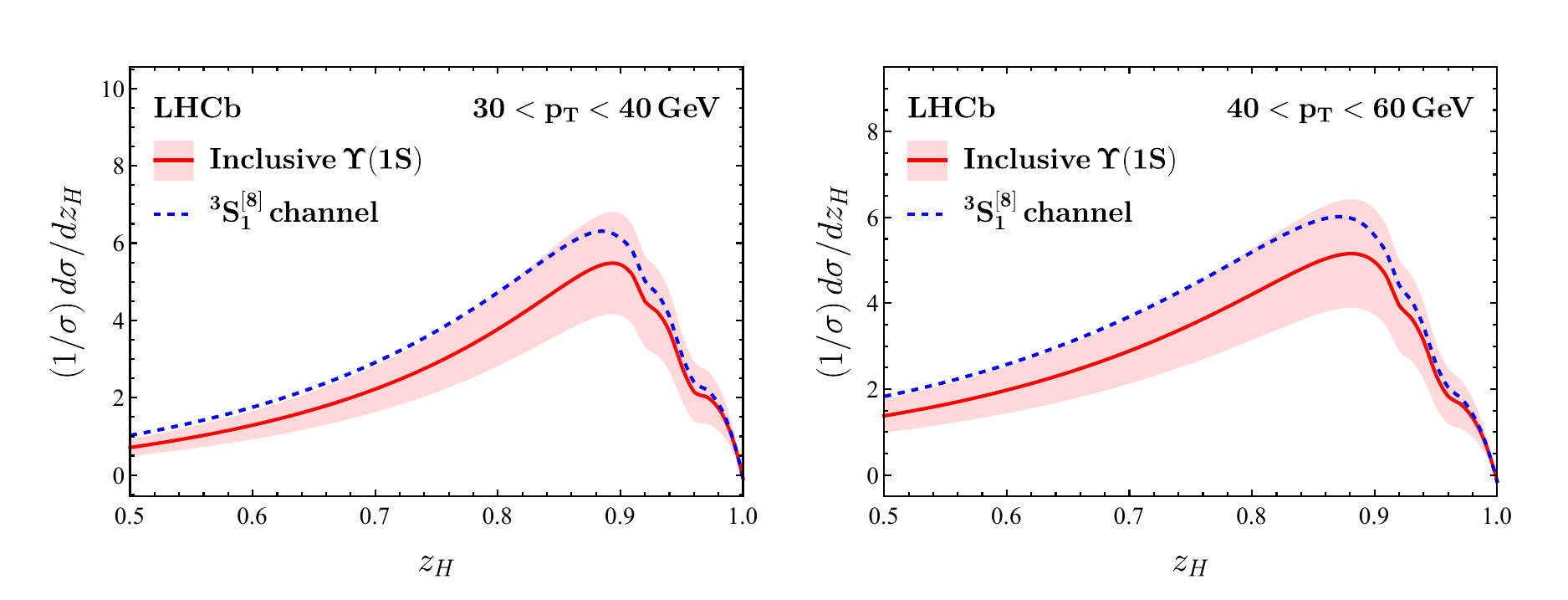}
\caption{
Normalized $z_H$ distributions for $\Upsilon(1S)$ production obtained with the Brambilla LDME set in LHCb kinematic regions:
(a) $30<p_T<40$ GeV and (b) $40<p_T<60$ GeV with $2.5<|\eta|<4$.
}
\label{fig:LHCbprediction}
\end{figure}

We provide predictions for LHCb in a complementary kinematic regime.
The corresponding LHCb muon acceptance requirements are implemented following the prescription described in Appendix~\ref{app:mucut}. \fig{LHCbprediction} shows the normalized $z_H$ distributions for $30<p_T<40$ GeV and $40<p_T<60$ GeV with $2.5<|\eta|<4$. The predictions are obtained for $\sqrt{s}=13$ TeV using anti-$k_T$ jets with $R=0.5$. While shown here for $\Upsilon(1S)$ and for the Brambilla set, this universal shape holds for other LDME sets and for $\Upsilon(2S, 3S)$, owing to the dominance of the ${}^3S_1^{[8]}$ channel.

Compared with the CMS kinematic region, the predictions exhibit a more pronounced large-$z_H$ enhancement, consistent with the expected $p_T$ dependence discussed above. At lower transverse momenta, the effect of fragmentation evolution is reduced, leading to a less broadened spectrum and a more localized large-$z_H$ peak.

\subsection{Dependence on jet transverse momentum and radius}
\label{sec:4-5}

We analyze the unnormalized $z_H$ spectra by varying $p_T$ while holding the quarkonium momentum $p_T^H$ fixed, highlighting both the $p_T$ dependence and the absolute predictions of the various LDME sets. In addition, we investigate jet-radius dependence at fixed $p_T$.

\begin{figure*}[tb]
\centering
\includegraphics[width=\textwidth]{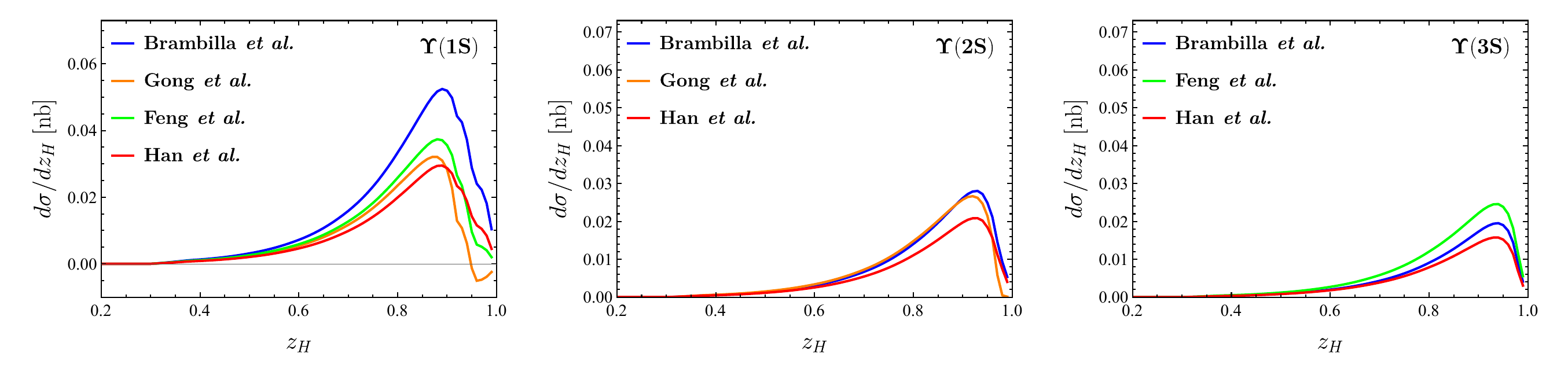}
\caption{
Inclusive $\Upsilon(1S)$ (left), $\Upsilon(2S)$ (center), and $\Upsilon(3S)$ (right) momentum-fraction distributions in the $p_T^H$ bin $60 < p_T^H < 75$ GeV. Predictions are shown without normalization.
}
\label{fig:qptint_inclusive}
\end{figure*}

\fig{qptint_inclusive} shows the inclusive $\Upsilon(1S)$, $\Upsilon(2S)$, and $\Upsilon(3S)$ momentum-fraction distributions in a fixed $p_T^H$ bin of $60 < p_T^H < 75$ GeV, with $|y^H|<1.5$. Unlike in previous sections where the quarkonium $p_T^H$ was varied, here $z_H = p_T^H/p_T$ changes as a function of the jet transverse momentum $p_T$. In addition, the results are shown without normalization to illustrate the absolute differences among the various LDME sets.

Several observations can be made from \fig{qptint_inclusive}. First, all LDME sets predict very similar distribution shapes despite noticeable differences in the overall normalization. 
The distributions are enhanced toward large $z_H$, since for fixed $p_T^H$ larger $z_H$ corresponds to smaller jet transverse momentum, where the production rate is larger. This enhancement is eventually suppressed as $z_H\to1$, where threshold effects become important.
Compared with the distributions discussed in Sec.~\ref{sec:4-3}, \fig{qptint_inclusive} shows a more pronounced suppression toward the low-$z_H$ region, reflecting the decrease of the production rate toward larger jet transverse momentum.

Second, the relative ordering of the LDME predictions varies among the three bottomonium states. For $\Upsilon(1S)$ production, the Brambilla LDME set yields the largest cross section over most of the $z_H$ range. For the excited states, the differences among the LDME sets are more visible in the absolute production rates. In particular, the Gong and Brambilla sets give comparable predictions for $\Upsilon(2S)$ production, while the Feng set provides the largest contribution for $\Upsilon(3S)$ production near the peak region. These differences originate primarily from the different hierarchy of color-octet matrix elements among the LDME sets.

\begin{figure}[tb]
\centering
\includegraphics[width=0.5\textwidth]{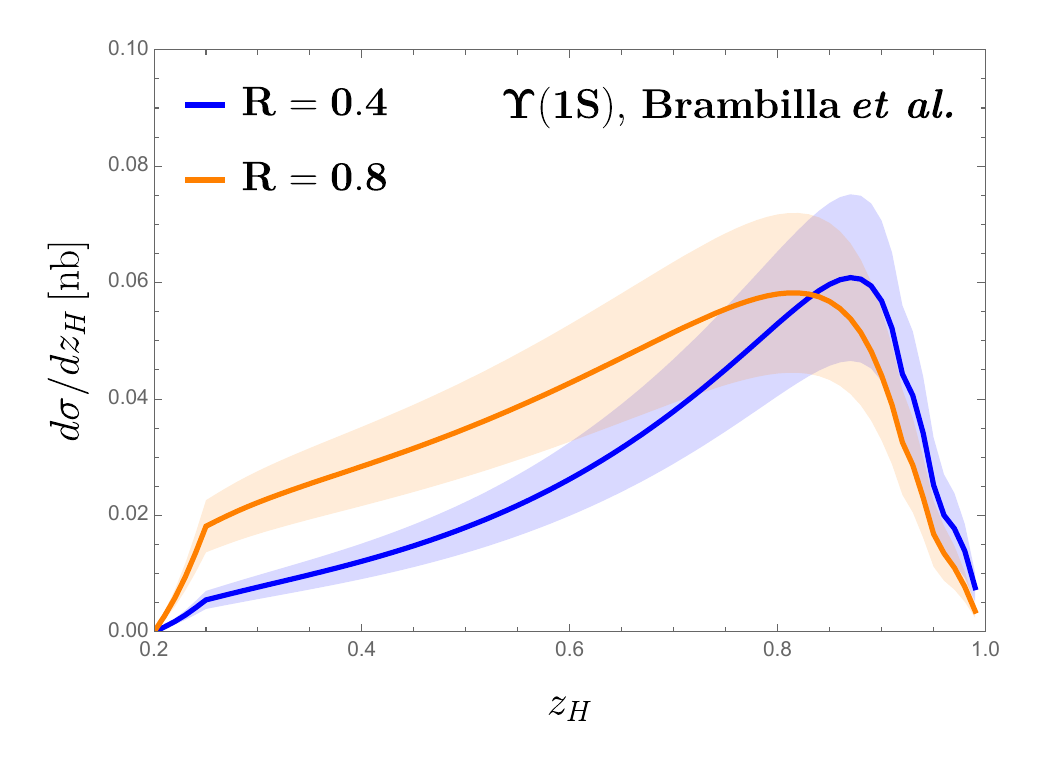}
\caption{
Dependence of the $\Upsilon(1S)$ $z_H$ spectrum on the jet radius, obtained with the Brambilla LDME set for $60<p_T<75$ GeV.
}
\label{fig:Rdependence}
\end{figure}

We next investigate the dependence of the $z_H$ distributions on the jet radius parameter $R$. \fig{Rdependence} compares predictions obtained with two representative jet radii, $R=0.4$ and $R=0.8$, using the Brambilla LDME set.

A clear and systematic dependence on the jet radius is observed. Reducing the jet radius from $R=0.8$ to $R=0.4$ enhances the threshold region and shifts the large-$z_H$ peak toward larger values of $z_H$, resulting in a more pronounced peak structure. At the same time, the spectrum is suppressed in the intermediate- and small-$z_H$ regions.

This behavior follows directly from the definition of the momentum fraction, $z_H = p_T^{H} / p_T$.
For a smaller jet radius, a larger fraction of the accompanying QCD radiation escapes the jet cone and therefore does not contribute to the reconstructed jet momentum.
As a result, the jet transverse momentum is systematically reduced, shifting the measured $z_H$ toward larger values. Conversely, for larger jet radii, more radiation is clustered into the jet, which increases the jet momentum and redistributes events toward smaller values of $z_H$. The jet-radius dependence therefore manifests itself primarily as a migration of event weight from the threshold region to intermediate and small momentum fractions as $R$ increases.

Although the jet-radius variation modifies the peak position and redistributes the event weight across the $z_H$ range, the overall structure of the distribution remains qualitatively unchanged. 
The sizable $R$ dependence indicates that quarkonium-in-jet observables are sensitive not only to the fragmentation dynamics but also to the details of the jet definition. Consequently, measurements using multiple jet radii could provide an additional probe of the underlying fragmentation dynamics and offer a useful test of the theoretical framework.

\section{Conclusion}
\label{sec:5}

We study bottomonium production inside jets as an extension of our previous analysis of charmonium, implementing NLO+LL threshold-resummed FFs within the FJF framework. By resumming the threshold logarithms arising in the FF, we obtain perturbatively reliable predictions in the large-$z_H$ region, where fixed-order calculations develop singular threshold-enhanced structures. 

We investigated the dependence of the predicted spectra on four representative bottomonium LDME sets, which exhibit substantial differences in the relative sizes and signs of the color-octet matrix elements.
Unlike the charmonium case, where the distributions exhibit significant sensitivity to different LDME sets, bottomonium distributions show only limited dependence on the choice of LDME set and display universal shapes among the different $\Upsilon(nS)$ states. We find that this feature stems from the dominance of the ${}^3S_1^{[8]}$ fragmentation channel. While the $^3S_1^{[8]}$ dominance does not strictly apply to the direct contribution, the large feeddown contributions from $\chi_b$ states are dominated by the ${}^3S_1^{[8]}$ channel, thereby reinforcing the universal shape of the inclusive $\Upsilon(nS)$ distributions. The universal shape is characterized by a peak structure in the large-$z_H$ region primarily controlled by the ${}^3S_1^{[8]}$ contribution, while DGLAP evolution reduces the peak and redistributes the spectrum toward smaller $z_H$ at higher jet transverse momentum $p_T$.

The characteristic interplay between the ${}^3S_1^{[8]}$ and ${}^3P_J^{[8]}$ channels, which provides significant discrimination among LDME scenarios in charmonium production, becomes less visible in bottomonium production inside jets.

We compared our predictions for  $\Upsilon(1S)$, $\Upsilon(2S)$, and $\Upsilon(3S)$ production with the recent CMS measurement of the momentum-fraction distributions. Our results reproduce the main features of the measured distributions, including the characteristic enhancement at large $z_H$. We also reproduce the systematic broadening of the distributions toward smaller $z_H$ with increasing jet transverse momentum. The overall agreement demonstrates the validity of the resummed FJF framework for bottomonium production inside jets.
Beyond the comparison with the CMS measurement, we also provide predictions for the LHCb experiment and for jet transverse-momentum and radius dependence.

The approximate universality relations for the Brambilla LDME set provide a possible interpretation of the similar $z_H$ shapes among the different $\Upsilon(nS)$ states. The relations imply that the relative sizes of the color-octet LDMEs are approximately independent of the radial excitation, while the remaining differences in the inclusive $\Upsilon(nS)$ distributions arise mainly from their different feeddown contributions. In addition, the dominance of the ${}^3S_1^{[8]}$ channel in the relevant $\chi_b$ feeddown contributions leads to similar fragmentation patterns and further reduces the sensitivity of the distributions to the radial excitation.

The CMS measurement also includes observables sensitive to the transverse momentum of the quarkonium relative to the jet axis. Extending the present study to the TMD FJF framework would provide a complementary probe of bottomonium production mechanisms and may offer additional sensitivity to the interplay between fragmentation dynamics and LDMEs. Another interesting direction is to incorporate finite heavy-quark mass effects into the FJF formalism. Such corrections are expected to become increasingly relevant at lower transverse momenta and may improve the description of bottomonium production in kinematic regions where the massless approximation becomes less accurate. We leave these investigations to future work.

\acknowledgments 
The work of DKK was supported by the faculty research fund of Sejong University in 2026.
The work of HSC was supported by the National Research Foundation of Korea (NRF) grant funded by the Korean government (MSIT) in 2026 (Grant No. RS-2026-25497906, Project Title: Hadron Production at Next-Generation Colliders). 
The work of YLW was supported by the Scientific Research Foundation of Chengdu University of Technology under Grant No. 10912-KYQD2026-11957.
We thank the Erwin-Schrödinger International Institute for Mathematics and Physics at the University of Vienna for partial support during the Program `New Paradigms for Harnessing Quantum Field Theory at Colliders', July 27 – August 28, 2026.

\appendix

\section{Cross section formula}
\label{app:factorization}

\subsection*{Transformation from $(v,w)$ to $(x_a,x_b)$}

Starting from Eq.~(8) of Ref.~\cite{Jager:2002xm}, the partonic variables are related to the hadronic momentum fractions by

\begin{align}
x_a &= \frac{VW}{v w z_c}, &
x_b &= \frac{1-V}{z_c(1-v)},
\end{align}
or equivalently,

\begin{align}
v &= 1-\frac{1-V}{z_cx_b},
&
w &= \frac{VW}{vz_cx_a}.
\end{align}

The Jacobian of the transformation is

\begin{align}
\left|
\frac{\partial(v,w)}
{\partial(x_a,x_b)}
\right|
=
\frac{VW(1-V)}
{z_c^2v\,x_a^2x_b^2},
\end{align}
which gives

\begin{align}
dv\,dw
=
\frac{VW(1-V)}
{z_c^2v\,x_a^2x_b^2}
dx_a\,dx_b .
\end{align}

Using

\begin{align}
1-v=\frac{1-V}{z_cx_b},
\qquad
w=\frac{VW}{vz_cx_a},
\end{align}
the integration measure becomes

\begin{align}
\frac{dv}{v(1-v)}
\frac{dw}{w}
=
\frac{dx_a\,dx_b}
{vx_ax_b}.
\end{align}

Therefore,

\begin{align}
E_H\frac{d\sigma}{d^3p_H}
=
\frac{1}{\pi S}
\sum_{abc}
\int_{z_0}^1
\frac{dz_c}{z_c^2}
\int
\frac{dx_a\,dx_b}
{v x_a x_b}
f_a(x_a)
f_b(x_b)
D_c^H(z_c)
\,
\frac{d\hat{\sigma}_{ab}^c}{dv\,dw}.
\end{align}

The partonic center-of-mass energy is related to the hadronic one through
\begin{align}
\hat{s}=x_a x_b S .
\end{align}

\subsection*{Transformation to $(p_T,\eta)$}

In the massless approximation,

\begin{align}
d^3p
=
E_H\,p_T\,dp_T\,d\eta\,d\phi ,
\end{align}
and after integrating over the azimuthal angle,

\begin{align}
E_H
\frac{d\sigma}{d^3p_H}
=
\frac{1}{2\pi}
\frac{d\sigma}
{p_T dp_T d\eta}.
\end{align}

At this stage, the partonic hard-scattering cross section appearing in the previous expression is differential in the partonic variables $(v,w)$. The corresponding differential cross section in $(p_T,\eta)$ is obtained through the change of variables
\begin{align}
\frac{d\hat{\sigma}_{ab}^c}{dp_T\,d\eta}
=
\frac{d\hat{\sigma}_{ab}^c}{dv\,dw}
\left|
\frac{\partial(v,w)}
{\partial(p_T,\eta)}
\right|.
\end{align}

For fixed $x_a$, $x_b$, and $z_c$, the relations between the two sets of variables can be written as
\begin{align}
v
&=
1-\frac{p_T e^{-\eta}}
{z_c\sqrt{x_a x_b S}},
\\
w
&=
\frac{p_T^2}
{z_c^2 x_a x_b S\,v(1-v)}.
\end{align}

The corresponding Jacobian is
\begin{align}
\left|
\frac{\partial(v,w)}
{\partial(p_T,\eta)}
\right|
=
\frac{2p_T}
{x_a x_b S\,z_c^2v}.
\end{align}

Therefore,
\begin{align}
\frac{d\hat{\sigma}_{ab}^c}{dv\,dw}
=
\frac{x_a x_b S\,z_c^2v}
{2p_T}
\frac{d\hat{\sigma}_{ab}^c}{dp_T\,d\eta}.
\end{align}

Substituting this relation into the previous expression, and using
\begin{align}
\frac{d\sigma}{dp_T d\eta}
=
2\pi p_T
E_H\frac{d\sigma}{d^3p_H},
\end{align}
we obtain an overall factor
\begin{align}
\frac{2p_T}{S}
\frac{1}{z_c^2v x_a x_b}
\end{align}
multiplying the inverse Jacobian factor
\begin{align}
\frac{x_a x_b S z_c^2v}{2p_T}.
\end{align}
These factors cancel exactly, leaving
\begin{align}
\frac{d\sigma}{dp_Td\eta}
=
\sum_{a,b,c}
\int
dz_c
\int
dx_a dx_b\,
f_a(x_a)
f_b(x_b)
D_c^H(z_c)
\frac{d\hat{\sigma}_{ab}^c}
{dp_Td\eta}.
\end{align}

\subsection*{Connection to the semi-inclusive FJF formalism}

In the semi-inclusive FJF formalism, the inclusive FF is replaced by a semi-inclusive FJF that describes the production of a hadron inside a reconstructed jet. The relevant momentum fractions are related by

\begin{align}
z_c
=
\frac{p_T^H}{p_T^i}
=
\frac{p_T}{p_T^i}
\frac{p_T^H}{p_T}
=
z z_H ,
\end{align}
where $p_T$ denotes the transverse momentum of the reconstructed jet, and

\begin{align}
z=\frac{p_T}{p_T^i},
\qquad
z_H=\frac{p_T^H}{p_T}.
\end{align}

At fixed $z_H$, the fragmentation variable can equivalently be parameterized by the jet momentum fraction,
\begin{align}
z=\frac{z_c}{z_H},
\qquad
dz_c=z_H\,dz.
\end{align}
The fragmentation contribution is then reorganized in terms of the two momentum fractions $z$ and $z_H$ and expressed through the semi-inclusive fragmenting jet function,
\begin{align}
D_c^H(z_c)
\longrightarrow
\mathcal{G}_i^H(z,z_H,p_TR,\mu),
\end{align}
whose definition incorporates the corresponding momentum-fraction normalization and Jacobian. This gives the standard $dz/z$ convolution measure in the semi-inclusive FJF factorization formula.

Consequently, the factorized cross section differential in $z_H$ becomes

\begin{align}
\frac{d\sigma}
{dp_T d\eta dz_H}
=
\sum_i
\int_{z_{\min}}^1
\frac{dz}{z}
\,
\frac{d\sigma^{i}}
{dp_T^i d\eta}
(p_T^i,\eta,\mu)
\,
\mathcal{G}_i^H(z,z_H,p_TR,\mu),
\end{align}
where the PDF convolution is included in the definition of the inclusive
parton production cross section,

\begin{align}
\frac{d\sigma^{i}}
     {dp_T^i d\eta}
(p_T^i,\eta,\mu)
&=
\sum_{a,b}
\int dx_a dx_b\,
f_{a/p}(x_a,\mu)
f_{b/p}(x_b,\mu)
\frac{d\hat{\sigma}_{ab\to i+X}}
     {dp_T^i d\eta}
(p_T^i,\eta,\mu).
\end{align}

This expression reproduces Eqs.~\eqref{eq:hadronic_sigma} and
\eqref{eq:parton_production} of the main text.

\section{Impact of threshold resummation}
\label{app:FOcomparison}

\begin{figure*}[t]
\centering
\includegraphics[width=0.9\textwidth]{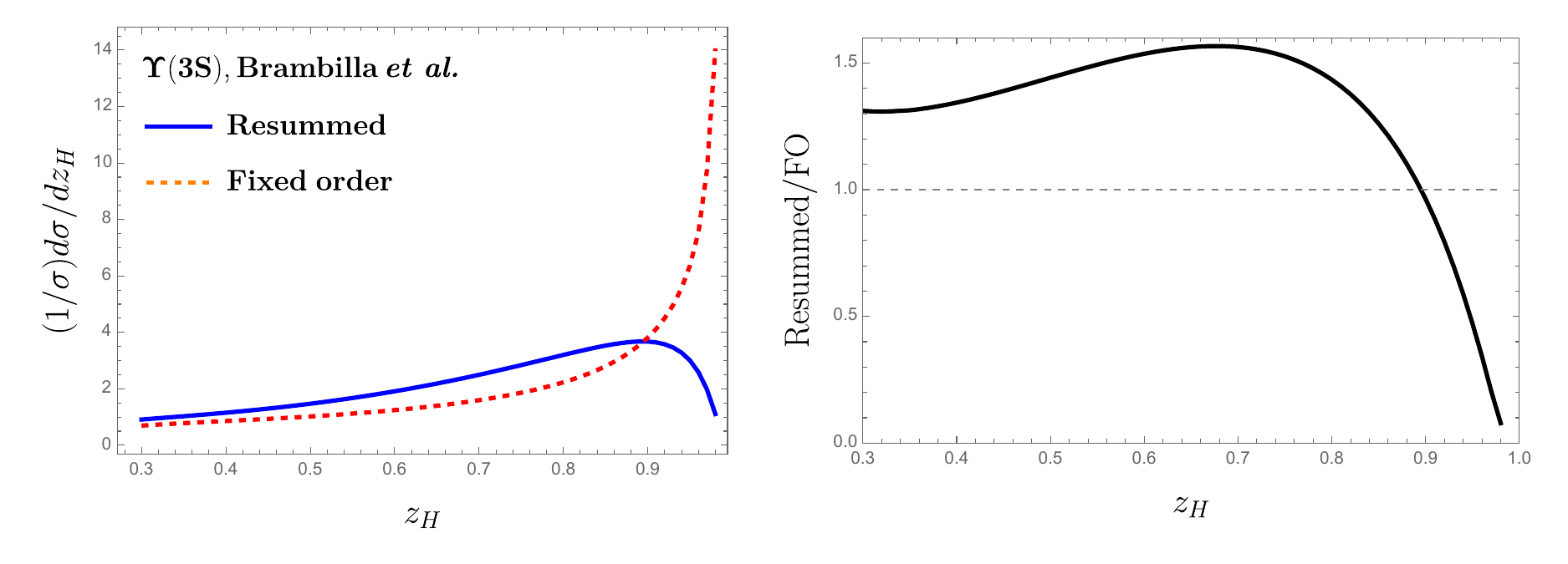}
\caption{
Comparison between the threshold-resummed and fixed-order predictions for the normalized  $\Upsilon(3S)$ $z_H$ distributions in the jet transverse-momentum interval $75<p_T<100$ GeV using the Brambilla LDME set. The left panels show the normalized spectra, while the right panels display the ratio of the resummed prediction to the fixed-order result.
}
\label{fig:FOcomparison}
\end{figure*}

To illustrate the numerical impact of threshold resummation, we compare the resummed prediction with the corresponding fixed-order calculation for $\Upsilon(3S)$ production. \fig{FOcomparison} shows the comparison for the intermediate jet transverse-momentum interval, $75<p_T<100$ GeV. Both predictions are normalized using the same normalization factor obtained from the resummed spectrum over the region $z_H>z_{\rm cut}$, excluding the first two experimental bins, in order to facilitate a direct comparison of the distribution shapes.

The ratio between the resummed and fixed-order predictions exhibits a nontrivial structure as a function of $z_H$. In the intermediate region away from the threshold, the resummed result is typically larger than the fixed-order prediction by approximately $20$--$30\%$, reaching up to about $50\%$ in some bins. This indicates the presence of sizeable higher-order logarithmic corrections, which remain numerically important but do not signal a breakdown of perturbative stability.

As $z_H \to 1$, the fixed-order prediction develops a pronounced threshold-enhanced structure due to large threshold logarithms, which become increasingly important and require resummation for a reliable prediction in the threshold region. Once resummation is included, these contributions are exponentiated and redistributed over multiple emissions, leading to a smooth suppression in the threshold region and a regularized behavior of the spectrum.

Overall, the comparison shows that threshold resummation mainly affects the threshold region, while producing moderate but controlled corrections over most of the phenomenologically relevant $z_H$ range.

\section{LDME sets}
\label{app:LDME}

\begin{table*}[ht!]
\centering

\footnotesize
\setlength{\tabcolsep}{4pt}
\renewcommand{\arraystretch}{1.15}

\begin{tabular}{|c|c|c|c|c|c|c|}
\hline
\multirow{2}{*}{\textbf{State}}
&
\multirow{2}{*}{\textbf{Category}}
&
\multirow{2}{*}{\textbf{Reference}}
&
$\langle O(^3S_1^{[1]}) \rangle$
&
$\langle O(^3S_1^{[8]}) \rangle$
&
$\langle O(^1S_0^{[8]}) \rangle$
&
$\langle O(^3P_0^{[8]}) \rangle /m_b^2$
\\

&
&
&
(GeV$^3$)
&
($10^{-2}$ GeV$^3$)
&
($10^{-2}$ GeV$^3$)
&
($10^{-2}$ GeV$^3$)
\\
\hline

\multirow{4}{*}{$\Upsilon(1S)$}

& 1
& Brambilla \textit{et al.}
& $9.28\pm0.93$
& $2.96\pm0.93$
& $-0.40\pm2.04$
& $2.12\pm0.68$
\\

& 2
& Gong \textit{et al.}
& 9.28
& $-0.76\pm0.24$
& $11.15\pm0.43$
& $-0.67$
\\

& 3
& Feng \textit{et al.}
& 9.28
& $0.47\pm0.41$
& $11.6\pm2.61$
& $-0.49\pm0.59$
\\

& 4
& Han \textit{et al.}
& 9.28
& $1.17\pm0.02$
& $13.7\pm1.1$
& 0
\\
\hline

\multirow{3}{*}{$\Upsilon(2S)$}
& 1
& Brambilla \textit{et al.}
& $4.96\pm0.50$
& $1.52\pm0.47$
& $-0.20\pm1.04$
& $1.08\pm0.35$
\\

& 3
& Gong \textit{et al.}
& 4.63
& $-0.01\pm0.82$
& $3.55\pm2.12$
& $-0.56\pm0.48$
\\

& 4
& Han \textit{et al.}
& 4.63
& $1.08\pm0.20$
& $6.07\pm1.08$
& 0
\\

\hline

\multirow{3}{*}{$\Upsilon(3S)$}
& 1
& Brambilla \textit{et al.}
& $3.83\pm0.38$
& $1.17\pm0.37$
& $-0.16\pm0.81$
& $0.84\pm0.27$
\\

& 3
& Feng \textit{et al.}
& 3.54
& $1.52\pm0.33$
& $-0.18\pm1.40$
& $-0.01\pm0.30$
\\

& 4
& Han \textit{et al.}
& 3.54
& $0.83\pm0.02$
& $2.83\pm0.07$
& 0
\\

\hline

\end{tabular}

\caption{
NRQCD LDMEs used for $\Upsilon(nS)$ production. All values are quoted at the common NRQCD factorization scale $\mu_\Lambda=m_b$.
}
\label{tab:SwaveLDMEs}

\end{table*}

\begin{table*}[ht!]
\centering

\footnotesize
\setlength{\tabcolsep}{4pt}
\renewcommand{\arraystretch}{1.15}

\begin{tabular}{|c|c|c|c|}
\hline
\multirow{2}{*}{\textbf{State}}
&
\multirow{2}{*}{\textbf{Reference}}
&
$\langle O(^3S_1^{[8]}) \rangle$
&
$\langle O(^3P_0^{[1]}) \rangle /m_b^2$
\\

&
&
($10^{-2}$ GeV$^3$)
&
($10^{-2}$ GeV$^3$)
\\
\hline

\multirow{4}{*}{$\chi_{b0}(1P)$}

& Brambilla \textit{et al.}
& $1.44 \pm 0.20$
& $1.55$
\\

& Gong \textit{et al.}
& $1.68 \pm 0.16$
& $1.51$
\\

& Feng \textit{et al.}
& $1.16 \pm 0.07$
& $1.51$
\\

& Han \textit{et al.}
& $0.73 \pm 0.09$
& $1.73$
\\

\hline

\multirow{4}{*}{$\chi_{b0}(2P)$}

& Brambilla \textit{et al.}
& $1.71 \pm 0.24$
& $1.86$
\\

& Gong \textit{et al.}
& $3.24 \pm 0.67$
& $1.73$
\\

& Feng \textit{et al.}
& $1.50 \pm 0.21$
& $1.73$
\\

& Han \textit{et al.}
& $1.08 \pm 0.14$
& $1.73$
\\

\hline

\multirow{3}{*}{$\chi_{b0}(3P)$}

& Brambilla \textit{et al.}
& $1.71 \pm 0.24$
& $1.86$
\\

& Feng \textit{et al.}
& $1.50 \pm 0.21$
& $1.73$
\\

& Han \textit{et al.}
& $1.08 \pm 0.14$
& $1.73$
\\

\hline

\end{tabular}

\caption{
NRQCD LDMEs used for $\chi_b(nP)$ production. All values are quoted at $\mu_\Lambda=m_b$.
}
\label{tab:PwaveLDMEs}

\end{table*}

Tables~\ref{tab:SwaveLDMEs} and \ref{tab:PwaveLDMEs} summarize the NRQCD LDME sets used in this work. The sets are taken from the Brambilla, Gong, Feng, and Han extractions~\cite{Brambilla:2022ayc,Brambilla:2021abf,Gong:2013qka,Feng:2015wka,Han:2014kxa}. 
All LDMEs listed in Tables~\ref{tab:SwaveLDMEs} and \ref{tab:PwaveLDMEs} are quoted at the common NRQCD factorization scale $\mu_\Lambda=\mu_f=m_b$. The Gong LDME set was originally extracted at the NRQCD scale $\mu_\Lambda=\mu_i=m_b v$. For consistency with the other LDME sets, we evolve the ${}^3S_1^{[8]}$ matrix elements to $\mu_\Lambda=\mu_f$ using the NRQCD renormalization-group evolution equations~\cite{Brambilla:2022ayc,Brambilla:2021abf} at fixed order.

For the $S$-wave bottomonium states, the evolution of the ${}^3S_1^{[8]}$ matrix elements is given by
\begin{align}
\langle O^{\Upsilon(nS)}({}^3S_1^{[8]})\rangle_{\mu_\Lambda=\mu_f}
=
\langle O^{\Upsilon(nS)}({}^3S_1^{[8]})\rangle_{\mu_\Lambda=\mu_i}
+
\frac{6(N_c^2-4)}{N_c m_b^2}
\frac{\alpha_s}{\pi}
\langle O^{\Upsilon(nS)}({}^3P_0^{[8]})\rangle
\ln\frac{\mu_f}{\mu_i}.
\end{align}

For the $P$-wave bottomonium states, the corresponding evolution of the ${}^3S_1^{[8]}$ matrix elements is given by
\begin{align}
\langle O^{\chi_{bJ}}({}^3S_1^{[8]})\rangle_{\mu_\Lambda=\mu_f}
=
\langle O^{\chi_{bJ}}({}^3S_1^{[8]})\rangle_{\mu_\Lambda=\mu_i}
+
\frac{4C_F}{3N_c m_b^2}
\frac{\alpha_s}{\pi}
\langle O^{\chi_{bJ}}({}^3P_J^{[1]})\rangle
\ln\frac{\mu_f}{\mu_i}.
\end{align}

The original values of $\langle O({}^3S_1^{[8]})\rangle$ at $\mu_\Lambda=m_b v$ are $(-0.41\pm0.24$,$0.30\pm0.78$,$1.27\pm0.16$,$2.76\pm0.67$)$\times10^{-2}\,\mathrm{GeV}^3$ for $\Upsilon(1S)$, $\Upsilon(2S)$, $\chi_{b0}(1P)$, and $\chi_{b0}(2P)$, respectively.
For the color-singlet S-wave states, we quote
$\langle O(^3S_1^{[1]})\rangle$ rather than
$\langle O(^3S_1^{[1]})\rangle/(2N_c)$.
For the P-wave states, heavy-quark spin symmetry implies
\[
\langle O^{\chi_{bJ}}(^3S_1^{[8]})\rangle =
(2J+1)
\langle O^{\chi_{b0}}(^3S_1^{[8]})\rangle ,
\]
and
\[
\langle O^{\chi_{bJ}}(^3P_J^{[1]})\rangle = 
(2J+1)
\langle O^{\chi_{b0}}(^3P_0^{[1]})\rangle .
\]
Therefore, only the $\chi_{b0}$ matrix elements are listed in Table~\ref{tab:PwaveLDMEs}.

\section{Muon acceptance effects}
\label{app:mucut}

\begin{figure}[t]
\centering
\begin{subfigure}{0.48\textwidth}
    \centering
    \includegraphics[width=0.9\textwidth]{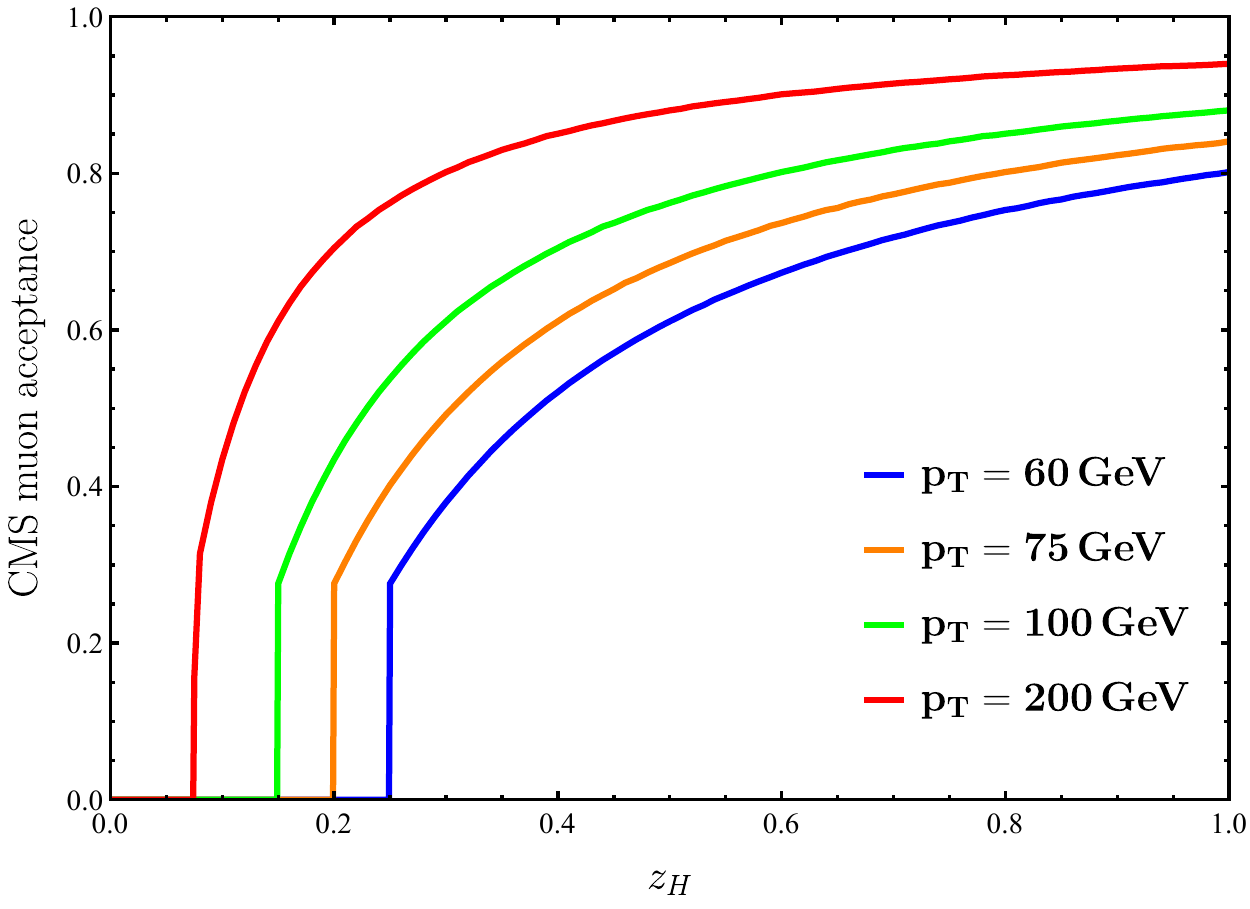}
    \caption{CMS acceptance}
    \label{fig:CMSmucut}
\end{subfigure}
\hfill
\begin{subfigure}{0.48\textwidth}
    \centering
    \includegraphics[width=0.9\textwidth]{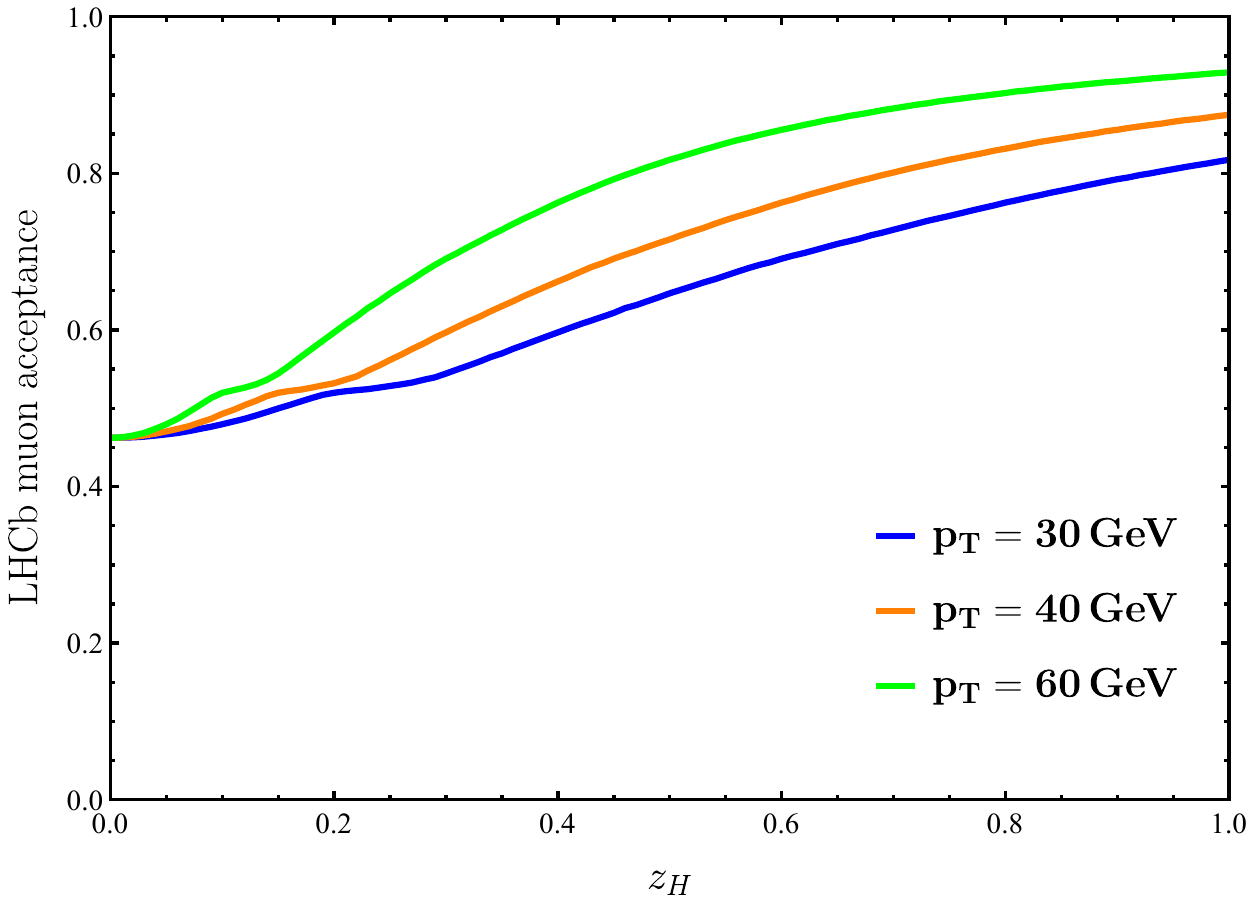}
    \caption{LHCb acceptance}
    \label{fig:LHCbmucut}
\end{subfigure}

\caption{
Muon acceptance effects on the $z_H$ distributions for CMS (left) and LHCb (right) kinematic selections. The acceptance is shown for representative jet transverse-momentum bins.
}
\label{fig:mucut}
\end{figure}

The impact of the muon acceptance requirements on the $z_H$ distributions is examined in \fig{mucut}. For the CMS selection~\cite{CMS:2026hnn}, we apply $p_T^\mu>6~{\rm GeV}$, $|\eta^\mu|<1.4$, $p_T^{\mu\mu}>15~{\rm GeV}$, and $|y^{\mu\mu}|<1.2$. For the LHCb selection, we adopt the requirements from Ref.~\cite{LHCb:2018yzj}: $p_T^\mu>1~{\rm GeV}$, $p^\mu>10~{\rm GeV}$, $2<y^\mu<4.5$, and $p_T^\mu p_T^{\bar{\mu}}>(1.3~{\rm GeV})^2$.

Both selections induce a suppression at low $z_H$, which originates from the muon reconstruction requirements rather than from the underlying quarkonium production dynamics. Since the muon cuts impose constraints on the quarkonium momentum, they translate into an effective lower bound on $z_H=p_T^H/p_T$ that depends on the jet transverse momentum.

The acceptance suppression is therefore weaker at higher jet transverse momentum, where the same muon requirements correspond to smaller values of $z_H$. The difference between the CMS and LHCb selections is also reflected in \fig{mucut}: due to the different muon kinematic requirements, the LHCb selection allows the distribution to extend further into the low-$z_H$ region.

This purely kinematic effect should be distinguished from the low-$z_H$ behavior arising from fragmentation evolution and feeddown contributions discussed in the main text.

\section{Distributions for $\Upsilon(2S)$ and $\Upsilon(3S)$}
\label{app:LDMEcomparison}

\begin{figure*}[htbp]
    \centering
    \begin{subfigure}{\textwidth}
        \centering
        \includegraphics[width=0.95\textwidth]{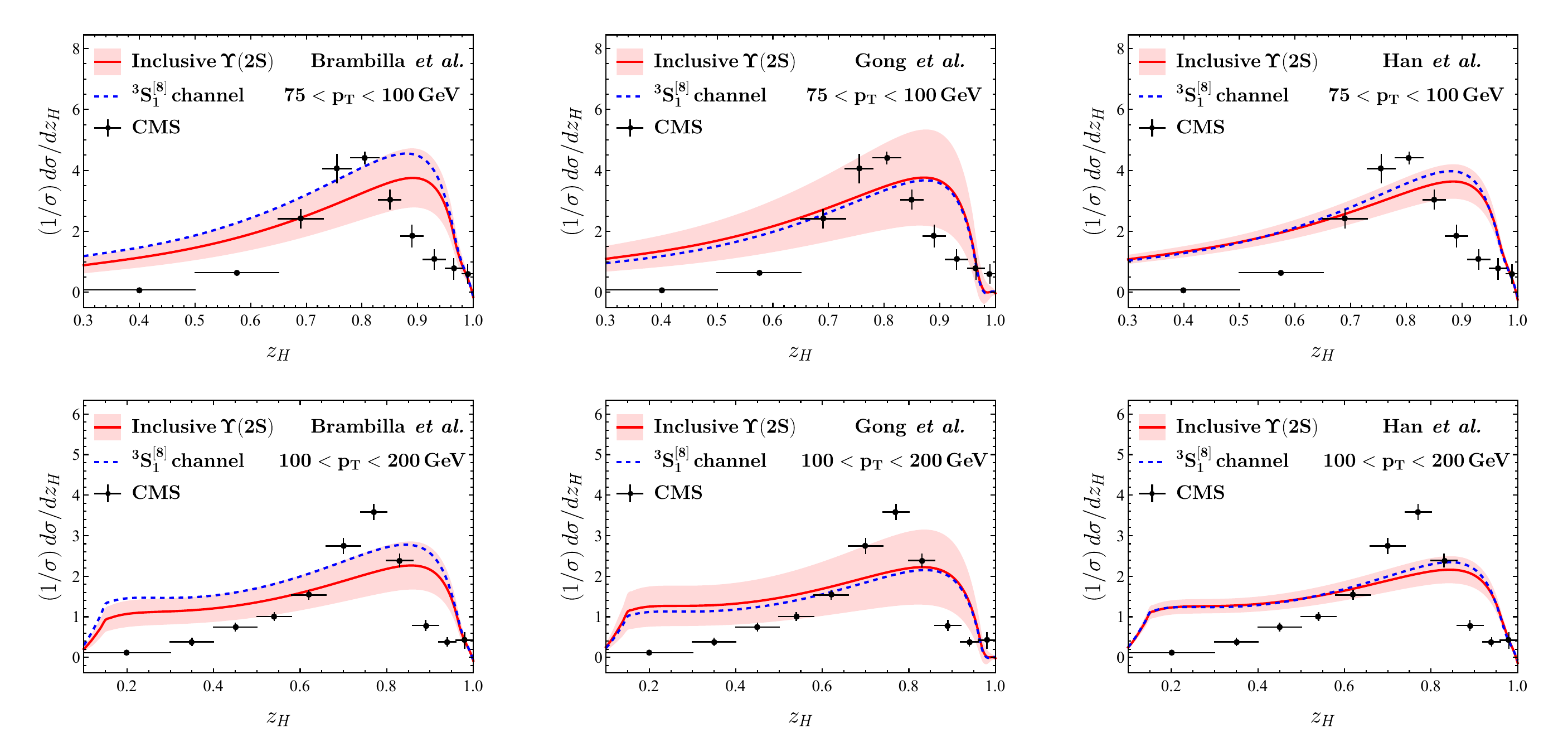}
        \caption{Inclusive $\Upsilon(2S)$ production}
        \label{fig:CMScomparison2S}
    \end{subfigure}
    
    \vspace{0.6cm}
    
    \begin{subfigure}{\textwidth}
        \centering
        \includegraphics[width=0.95\textwidth]{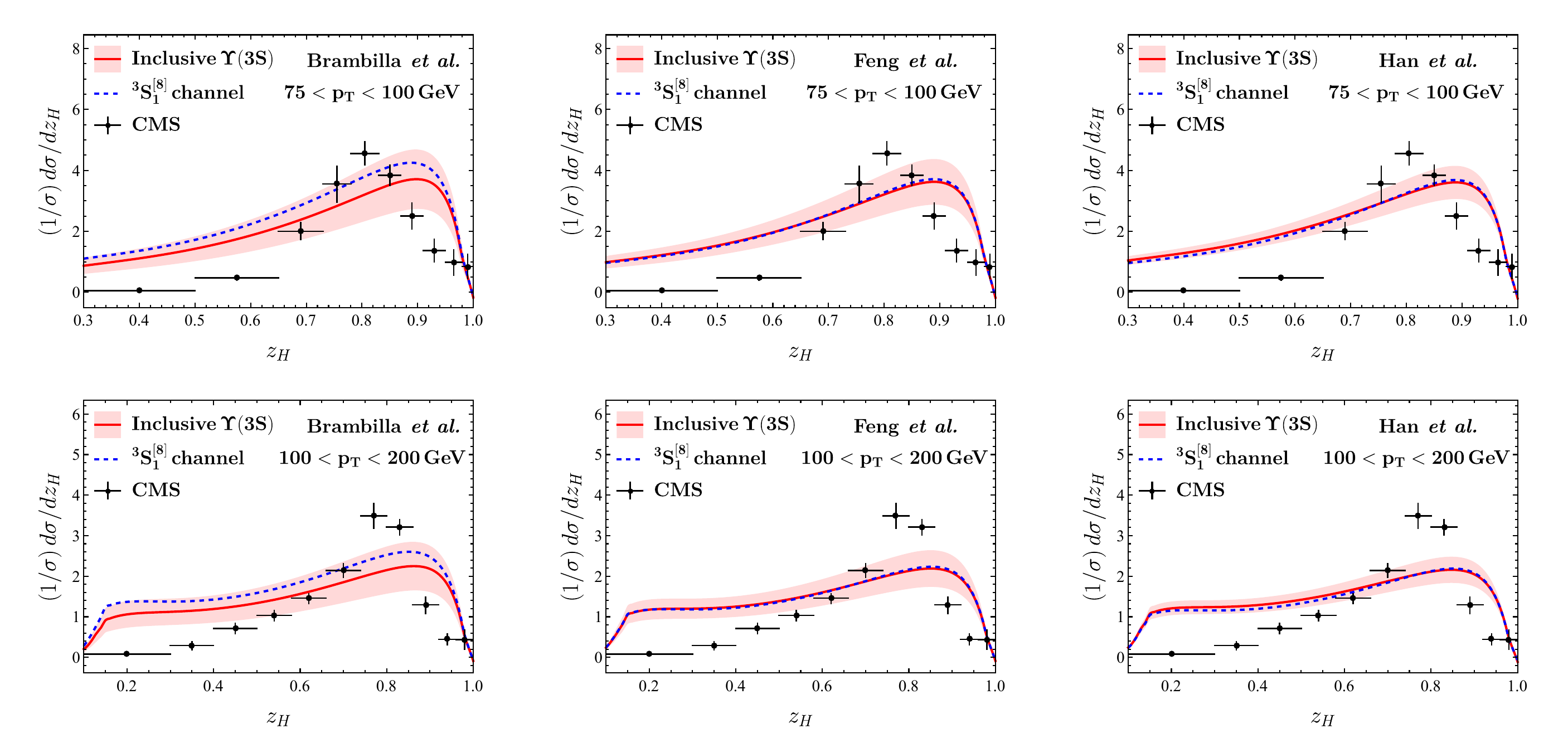}
        \caption{Inclusive $\Upsilon(3S)$ production}
        \label{fig:CMScomparison3S}
    \end{subfigure}
    
    \vspace{0.2cm}
    \caption{Dependence of the normalized $z_H$ distributions on the choice of LDME sets for (a) $\Upsilon(2S)$ and (b) $\Upsilon(3S)$ production inside jets. Results are shown for the two jet transverse-momentum intervals $75<p_T<100$ GeV, and $100<p_T<200$ GeV.}
    \label{fig:LDMEcomparisonExcited}
\end{figure*}

In the main text, the comparison with CMS data for $\Upsilon(2S)$ and $\Upsilon(3S)$ production is presented using the Brambilla LDME set as a representative choice. Since the normalized $z_H$ distributions of the excited bottomonium states exhibit limited sensitivity to the choice of LDMEs, the qualitative conclusions discussed in Sec.~\ref{sec:4-3} remain unchanged for the other available LDME sets.

For completeness, \fig{LDMEcomparisonExcited} shows these comparisons in the higher-$p_T$ bins, $75 < p_T < 100$ GeV and $100 < p_T < 200$ GeV, using the Brambilla, Gong, and Han LDME sets for $\Upsilon(2S)$ and the Brambilla, Feng, and Han sets for $\Upsilon(3S)$. Although moderate differences in the predicted spectra can be observed, particularly at low and intermediate $z_H$, all LDME sets lead to similar overall shapes and exhibit the same systematic trends with increasing jet transverse momentum.

\clearpage

\bibliographystyle{JHEP}
\bibliography{main.bbl}

\end{document}